\RequirePackage[2020-02-02]{latexrelease}
\documentclass[twocolumn, switch]{article} 

\usepackage{preprint}

\usepackage{amsmath, amsthm, amssymb, amsfonts}

\usepackage[numbers,square]{natbib}
\usepackage[utf8]{inputenc}	
\usepackage[T1]{fontenc}	
\usepackage{xcolor}		
\usepackage[colorlinks = true,
            linkcolor = purple,
            urlcolor  = blue,
            citecolor = cyan,
            anchorcolor = black]{hyperref}	
\usepackage{booktabs} 		
\usepackage{nicefrac}		
\usepackage{microtype}		
\usepackage{lineno}		
\usepackage{float}			
\usepackage{amsmath}
\usepackage{bm} 

\usepackage{lipsum}		

\usepackage{newfloat}
\DeclareFloatingEnvironment[name={Supplementary Figure}]{suppfigure}
\usepackage{sidecap}
\sidecaptionvpos{figure}{c}
\usepackage{pgfplots}
\usepackage{ragged2e}
\usepgfplotslibrary{groupplots}
\usepackage{subfigure}

\pgfplotsset{
	compat=newest,
	xlabel near ticks,
	ylabel near ticks
}

\usepackage{titlesec}
\titlespacing\section{0pt}{13pt plus 3pt minus 3pt}{1pt plus 1pt minus 1pt}
\titlespacing\subsection{0pt}{12pt plus 3pt minus 3pt}{1pt plus 1pt minus 1pt}
\titlespacing\subsubsection{0pt}{10pt plus 3pt minus 3pt}{1pt plus 1pt minus 1pt}

\usepackage{tikz,xcolor,hyperref}

\definecolor{lime}{HTML}{A6CE39}
\DeclareRobustCommand{\orcidicon}{
	\begin{tikzpicture}
	\draw[lime, fill=lime] (0,0) 
	circle [radius=0.16] 
	node[white] {{\fontfamily{qag}\selectfont \tiny ID}};
	\draw[white, fill=white] (-0.0625,0.095) 
	circle [radius=0.007];
	\end{tikzpicture}
	\hspace{-2mm}
}
\foreach \x in {A, ..., Z}{\expandafter\xdef\csname orcid\x\endcsname{\noexpand\href{https://orcid.org/\csname orcidauthor\x\endcsname}
			{\noexpand\orcidicon}}
}

\title{MAL2GCN: A Robust Malware Detection Approach Using Deep Graph Convolutional Networks With Non-Negative Weights}

\usepackage{xwatermark}

\usepackage{authblk}

\author[$ $]{Omid~Kargarnovin}
\author[$ $]{Amir~Mahdi~Sadeghzadeh}
\author[$ $]{Rasool~Jalili}

\affil[$ $]{Data and Network Security Lab, Department of Computer Enginnering, Sharif University of Technology, Tehran, Iran.}
\affil[$ $]{E-mail: kargarnovin@ce.sharif.edu; amsadeghzadeh@ce.sharif.edu;  jalili@sharif.edu.}

\begin{document}

\twocolumn[ 
  \begin{@twocolumnfalse} 
  
\maketitle

\begin{abstract}
With the growing pace of using Deep Learning (DL) to solve various problems, securing these models against adversaries has become one of the main concerns of researchers. Recent studies have shown that DL-based malware detectors are vulnerable to adversarial examples. An adversary can create carefully crafted adversarial example to evade DL-based malware detectors.
In this paper, we propose Mal2GCN, a robust malware detection model that uses Function Call Graph (FCG) representation of executable files combined with Graph Convolution Network (GCN) to detect Windows malware. 
Since FCG representation of executable files is more robust than raw byte sequence representation, numerous proposed adversarial example generating methods are ineffective in evading Mal2GCN. Moreover, we use the non-negative training method to transform Mal2GCN to a monotonically non-decreasing function; thereby, it becomes theoretically robust against appending attacks. We then present a black-box source code-based adversarial malware generation approach that can be used to evaluate the robustness of malware detection models against real-world adversaries. The proposed approach injects adversarial codes into the various locations of malware source codes to evade malware detection models. The experiments demonstrate that Mal2GCN with non-negative weights has high accuracy in detecting Windows malware, and it is also robust against adversarial attacks that add benign features to the Malware source code. 
\end{abstract}
\vspace{0.35cm}

  \end{@twocolumnfalse} 
] 



\section{Introduction}
\label{sec:introduction}
Despite their excellent performance of Deep Learning (DL) models in various tasks, it has been shown that adversarial examples can easily fool them. Adversarial examples are carefully-crafted inputs that cause the target model to misclassify them \cite{goodfellow2014explaining,akhtar2018threat,brendel2017decision,ilyas2018black,papernot2016limitations}. 
This vulnerability of deep learning models also exists in the cybersecurity domain, such as DL-based malware detection models. These malware detection models have been shown to be easily fooled by well-crafted adversarial manipulations to the malware binaries \cite{suciu2019exploring,hu2017generating,anderson2018learning,kolosnjaji2018adversarial,demetrio2021adversarial,maiorca2020adversarial,li2021irl}.\\
Defending against adversarial examples is an urgent task, and models must become robust against attacks before being deployed in the real world, especially in security-related fields such as malware detection. Recently, different approaches have been proposed to build robust malware detection models, such as using adversarial training \cite{al2018adversarial}, detecting adversarial examples \cite{alasmary2020soteria}, and the non-negative training method \cite{fleshman2018non}. However, recent studies have shown these approaches are still vulnerable to simple adversarial example generation methods, and there is still a need for a robust malware detection model that cannot be easily evaded by adversaries.\\
We propose Mal2GCN, a robust malware detection model that uses Function Call Graph (FCG) representation of executable files combined with Graph Convolution Network (GCN) to detect Windows malware. Recent studies have shown that the raw byte sequence of executable files is not a robust representation for detecting malware \cite{ceschin2019shallow}. This representation does not consider the functionality of the executables and only relies on the patterns in the byte sequence. Therefore, models that use raw byte sequence representation are simply evaded by adding benign patterns to the byte sequence of malware, such as appending attacks \cite{kolosnjaji2018adversarial,suciu2019exploring,demetrio2021functionality}. Mal2GCN uses FCG representation of executable files and combines it with GCN to better represent the functionality and relation between different modules of executable files than raw byte sequence. Hence, adding benign patterns to the byte sequence of malware without changing its functionality can not evade Mal2GCN; thereby, numerous proposed adversarial example generating methods are ineffective in evading Mal2GCN. Moreover, we use the non-negative training method \cite{fleshman2018non} to transform Mal2GCN to a monotonically non-decreasing function. Hence, Mal2GCN becomes theoretically robust against attacks that append or inject junk or benign codes into malware source code. To evaluate the robustness of Mal2GCN, we also present a black-box source code-based adversarial malware generation approach that injects adversarial codes into the malware source code. The generated adversarial malware can be used to evaluate the robustness of  malware detection models against more complex attacks. The main contributions of this work are as follows: 
\begin{itemize}
\item We argue and demonstrate that Function Call Graph (FCG) is more robust and accurate representation for malware detection than raw byte sequence representation.
\item We propose Mal2GCN, a robust and accurate malware detection model. Non-negative Mal2GCN is a monotonically non-decreasing function, which is theoretically robust against numerous adversarial malware attacks.
\item We present the first black-box adversarial malware generation approach based on injecting benign-looking codes into the malware source codes. 
\item We create a new comprehensive dataset for malware detection, which will be publicly available for researchers. 
\end{itemize}

The rest of the paper is organized as follows. In Sec. \ref{sec:Background}, graph convolutional networks and adversarial examples are explained. Sec. \ref{sec:Related} reviews previous studies on malware detection. It also reviews adversarial example attacks and defenses in the malware detection domain. Mal2GCN is introduced in Sec. \ref{sec:Mal2GCN}. In Sec. \ref{sec:Adversarial}, the adversarial source code generation approach is presented.
In Sec. \ref{sec:datasetandsetup}, the dataset is presented.
Sec. \ref{sec:Evaluation} evaluates the performance of Mal2GCN and its robustness against the adversarial source code generation attack. Sec. \ref{sec:Discussion} discusses the limitation of this study, and lastly, in Sec. \ref{sec:Conclusion}, this study is concluded with discussion on achievements and future research directions.

\section{Background}
\label{sec:Background}
The graph convolutional networks and adversarial examples are explained in this section.

\subsection{Graph Convolutional Networks (GCNs)}
\begin{figure}[h]
	\centering
	\includegraphics[width=0.5\textwidth]{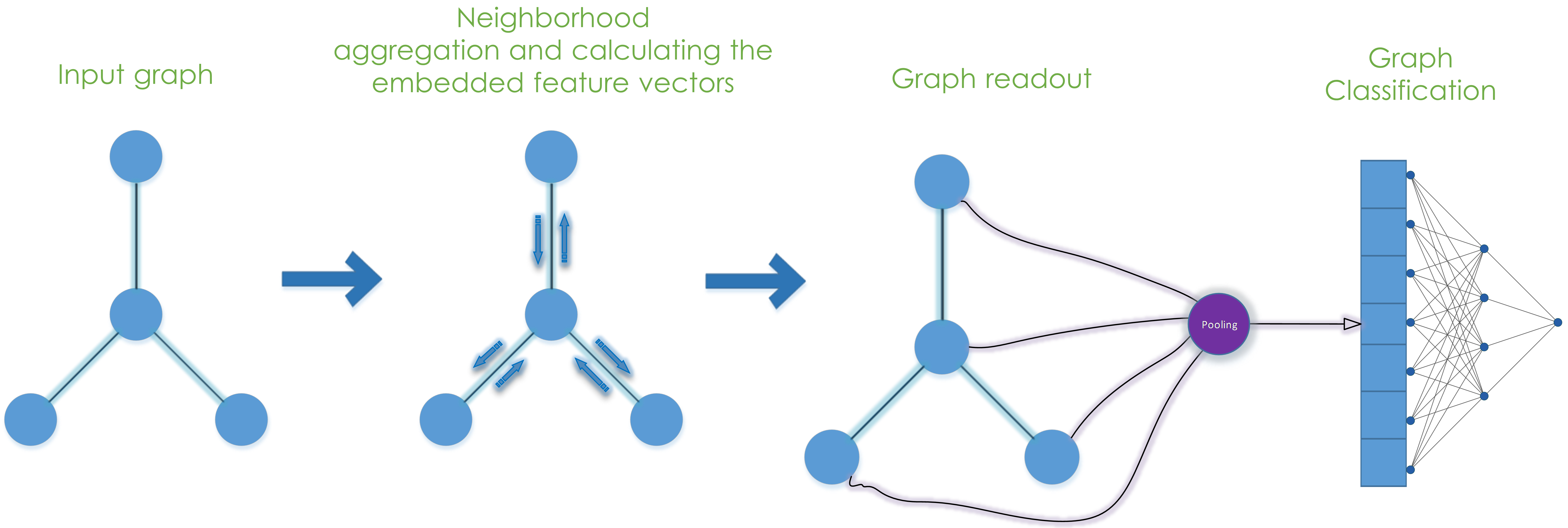}
	\caption{\label{fig:GCN} Classifying graphs using graph convolutional networks.}
\end{figure}

One of the most popular methods to use graph structures in machine learning is graph embedding. Graph embedding maps a graph into an embedded space in which the graph structural information and graph properties are maximally preserved. The amount of information that is preserved in the resulting embedded graph depends on how powerful the graph embedding algorithm is \cite{Errica2020fair}.
Graph Convolutional Network (GCN) is one of the approaches for graph embedding. GCN takes a graph as input in which each node has a corresponding feature vector that represents the features of that node and follows a recursive neighborhood aggregation where each node aggregates feature vectors of its neighbors to compute its embedded feature vector. After $k$ iterations of aggregation, a node's final embedded feature vector captures the structural information within that node's $k$-hop neighborhood. After the embedded feature vector of each node of the graph is calculated, a readout function, such as sum, average, or maximum, is applied on these vectors to calculate the feature vector of the entire graph \cite{xu2018how}. Lastly, we can use this feature vector for classifying the graph using a neural network model. Figure \ref{fig:GCN} shows an example of graph classification using GCN. A GCN can be formulated as follows:
\begin{equation}
H^{(l+1)} = \phi(\tilde{D}^{-\frac{1}{2}}\tilde{A}\tilde{D}^{-\frac{1}{2}}H^{(l)}W^{(l)})
\end{equation}
where $H^{(l)}$ is the feature vector for the $l^{th}$ layer of the neural network, $\phi$ is a non-linear function, $W^{(l)}$ is the weight matrix for the $l^{th}$ layer, and $D$ and $A$ represent the degree matrix and adjacency matrix, respectively. The $\tilde{D}$ and $\tilde{A}$ are the altered version of $D$ and $A$ in which a self-connection is added to each node of the graph. The shape of the input $H^{(0)}$ is $n \times d$, where n is the number of nodes in the graph and $d$ is the dimension of the initial feature vector for the nodes, which is dependent on the problem. If the GCN has $m$ layers, $H^{(m)}$ represents the final embedded feature vectors of nodes.\\

\subsection{Adversarial Examples}
An adversarial example is a malicious input that causes a machine learning model to make a false prediction \cite{goodfellow2014explaining,akhtar2018threat,brendel2017decision,ilyas2018black,papernot2016limitations}. Generally, an adversarial example is created by adding a small perturbation to a natural sample. Suppose that the label of natural sample $x$ is $y$, adversarial example $x'$ for victim classifier $f$ is defined as follows: 
\begin{equation}
\begin{split}
x' =  x + \xi \quad s.t. \quad f(x') = y',\; y\neq y',  \; x' \in domain(x)
\end{split}
\end{equation}
where $\xi$ is the perturbation. The proposed attacks to generate adversarial examples can be divided into \textit{black-box} and \textit{white-box} categories. In the white-box setting, the adversary has access to the model parameters; thereby, she can use the gradient of the victim model parameters to generate adversarial examples.
In the black-box setting, the adversary does not have access to the victim model parameters, and she only can send a query to the victim model and receive the prediction. Therefore, it is more difficult to conduct a successful attack in the black-box setting, but the attack is more realistic, considering it is very rare for attackers to have access to the malware detection model parameters.

\section{Related Work}
\label{sec:Related}
This section reviews the most prominent malware detection methods, adversarial example attacks, and defenses presented so far in the malware detection domain.

\subsection{Malware Detection}

So far, there have been various DL-based models proposed for malware detection, and these models depend on various features extracted from malware executables using static and/or dynamic analysis. One of the most popular DL-based malware detection models is MalConv \cite{raff2018malware}. MalConv takes the entire executable file as input and embeds each byte into an 8-dimensional vector. It then uses Convolutional Neural Networks (CNNs) to determine the probability that the executable file might be malware.
Using graph-based features to solve the problem of malware detection has also gained popularity in recent years \cite{Jiang2019,john2020graph,pei2020amalnet,yan2019classifying,wuchner2015robust,hashemi2017graph,nguyen2018auto,frenklach2021android,ou2022s3feature}. 
There are mainly three types of graph information that have been used by recent research for malware detection: function call graph (FCGs) \cite{Jiang2019,ou2022s3feature}, system call dependency graphs \cite{john2020graph}, and control flow graphs (CFGs) \cite{yan2019classifying,nguyen2018auto}. The function call graph is a directed graph in which each node represents a function in the program, and an edge from node $a$ to node $b$ corresponds to function $a$ calling function $b$ at some point. The system call dependency graph is a directed graph, in which each node represents a system call, and an edge from node $a$ to node $b$ corresponds to the data dependency between them, meaning the input of system call $a$ depends on system call $b$. 
The control flow graph is a directed graph that represents all the possible execution paths that can be taken during the program execution. Each node represents a basic block, and each directed edge corresponds to a possible control flow between the basic blocks.\\
In recent years, using Graph Neural Networks (GNNs) to embed graph structures into a feature vector has gained popularity \cite{xu2018how}. Researchers have also used GNNs to embed various types of graphs into a feature vector to solve the problem of malware detection or classification \cite{john2020graph,pei2020amalnet,yan2019classifying,schranko2019behavioral,gao2021gdroid}. Although graph neural networks can be used to learn powerful representations of graphs, they are still vulnerable to attacks, and using them alone will not make our malware detection model robust against attacks \cite{zugner2019adversarial}.

\subsection{Adversarial Malware Generation}
This section reviews the most prominent white-box and black-box approaches for generating adversarial examples against malware detectors presented in the previous studies.

\subsubsection{White-Box Attacks}

Grosse \textit{et al.}~\cite{grosse2017adversarial} proposed a gradient-based white-box attack against a malware classifier that takes the binary vector $ x \in  \{  0  ,  1  \}^m  $ as input that represent the corresponding malware, and the output is $ F(x) = [ F_{1} ,  F_{2} ]$, which $F_{1}$ and $F_{2}$ correspond to probability of input being a malware or benign executable, respectively. They used the gradient of the loss function with respect to the parameters of the DL-based classifier to generate adversarial malware by converting some of the 0's in the input feature vector to 1's. The main difference between their work and the previous adversarial generation methods was the restriction of not being able to remove features from the input because that will cause the malware to lose its original functionality. Al-Dujaili \textit{et al.}~\cite{al2018adversarial} improved this idea and proposed four new gradient-based methods to generate adversarial malware, called $dFGSM^k$, $rFGSM^k$, $BGA^k$, and $BCA^k$, which outperformed the previous study.
Kolosnjaji \textit{et al.} \cite{kolosnjaji2018adversarial} proposed a white-box attack against the MalConv model. They append adversarial bytes that were calculated using the gradient to different parts of the PE file without causing the malware to lose its original functionality.
Kruek \textit{et al.} \cite{kreuk2018deceiving} extend the work of Kolosnjaji \textit{et al.} and propose a method for constructing the malware sample, given the adversarial example embedding. 

\subsubsection{Black-Box Attacks}

Rigaki \textit{et al.}~\cite{rigaki2018bringing} presented a method to modify a malware to generate adversarial network traffic that causes a machine learning-based IPS to misclassify the malicious traffic as benign. They used Generative Adversarial Networks to convert malicious $C\&C$ network traffic to benign-looking traffic such as network traffic of Facebook.
Hu \textit{et al.}~\cite{hu2017generating} used a Generative Adversarial Network to convert the malware's feature vector to a feature vectors similar to benign executables. The authors represented each malware executable as a binary feature vector associated with the malware's Import Address Table (IAT). They then used the trained generator to add new adversarial functions to IAT in order to bypass an IAT-based malware detection model. As explained by Kawai \textit{et al.} \cite{kawai2019improved}, the main drawbacks of~\cite{hu2017generating} is that they use the same feature quantities for learning malware detection as they do for generating adversarial malware. They also used multiple malware instead of one, which affects the performance of avoidance. To resolve these issues, Kawai \textit{et al.} used differentiated learning methods with the different feature quantities and only used one malware to generate adversarial malware.
Anderson \textit{et al.} \cite{anderson2018learning} used reinforcement learning to generate adversarial malware. They trained an agent to find optimal modification to malware that can cause the target malware detection model to misclassify different malwares as benign. Vaya \textit{et al.} \cite{pesidious} improved this idea and combined the reinforcement learning with the generative adversarial network proposed by Hu \textit{et al.}~\cite{hu2017generating} to improve the success rate of attack. 
Demetrio \textit{et al.} \cite{demetrio2021functionality} proposed a genetic programming approach to evade static malware detectors in the black-box setting. The proposed attack is more query efficient compared to their previous works, and it guarantees to preserve the functionality of the original executable. They also added a penalty for the size of the injected adversarial payload to reduce the size of perturbation.
Abusnaina \textit{et al.} \cite{abusnaina2019adversarial} proposed a method to generate adversarial malware against a graph-based IoT malware detection model. This model extracts 23 features from the corresponding graph, such as the number of nodes, and uses them as the input to the model to detect malware. To attack this model, they selected six different graphs from the benign and malicious samples based on the graph size and then combined them to generate adversarial examples.

\subsection{Defending Against Adversarial Malware Attacks}

Following the emergence of adversarial examples, various defenses have been proposed.
One of the most popular approaches to make models robust against adversarial examples is adversarial training. In this approach, adversarial examples are used during training to make the model robust against them. 
Al-Dujaili \textit{et al.} \cite{al2018adversarial} proposed four different white-box methods to generate adversarial malware, then showed that by doing adversarial training with each method, the model will become robust against that specific method of adversarial malware generation. Zhang \textit{et al.} \cite{zhang2018the} and Sadeghzadeh \textit{et al.} \cite{9408630} emphasized the limitations of adversarial training and showed that adversarially trained models can still remain vulnerable to adversarial examples. Many researchers in recent years have used adversarial training to make their malware detection models robust \cite{rathore2020robust, khoda2019robust,wu2018enhancing,chen2019adversarial,li2020enhancing}, but as showed by \cite{al2018adversarial,zhang2018the}, these models are still vulnerable to adversarial examples that are generated using novel approaches.\\
One other similar defense method used in some of the recent studies is training a separate model to detect adversarial examples \cite{alasmary2020soteria}. In this method, the input of the model is first given to the adversarial example detection model. If the input is detected as an adversarial example, it will be discarded and will receive a malicious label; otherwise, the input will be given to the main model for detection. However, similar to the adversarial training method, if the attacker changes the adversarial generation method and uses novel approaches to generate adversarial examples, the model will no longer be able to detect them.\\
Demontis \textit{et al.} \cite{demontis2017yes} first showed the vulnerabilities of an Android based malware detector, called Drebin \cite{arp2014drebin}, against adversarial examples. They then proposed a robust training approach, which its underlying idea is to enforce the classifier to learn a evenly distributed feature weights, therefore the attackers will no longer be able to bypass the models with small and simple modifications.
Lastly, Fleshman \textit{et al.} \cite{fleshman2018non} proposed a novel approach to defend against adversarial example attacks, and they trained a robust model called Non-Negative Weight MalConv. They showed their model is resistant against recently proposed attacks, such as the appending attack proposed by \cite{kolosnjaji2018adversarial}. To make a binary classifier such as a malware detection model robust against adversarial example attacks, they proposed to restrict the model only to learn non-negative weights. This will cause the model only to predict based on the parts of the input feature vector that cause the output of the model to go towards label 1 (malware). Nevertheless, other studies \cite{ceschin2019shallow,Ebrahimi2020Binary,10.1145/3433210.3453086} show that non-negative MalConv model is still vulnerable to appending attacks.
\section{Mal2GCN}
\label{sec:Mal2GCN}
We propose Mal2GCN, a robust and accurate malware detection approach using deep graph convolutional networks combined with non-negative weights training method. Mal2GCN employs the function call graphs of executables as input to the GCN and uses the API calls and the referenced strings inside each function to calculate the feature vector for each node (function). It uses a two-layer \textit{Graph Convolutional Network (GCN)}, combined with a \textit{Graph Classifier (GClf)} to detect malware. All steps required for classifying an executable using Mal2GCN are explained in the following.

\subsection{API and Strings Extraction}

With the help of IDA pro \cite{IDA}, Mal2GCN extracts all the call instructions in each function and detects whether the destination of a call instruction is a library function or not. The library function could either be a local function, which happens when the library is statically linked to the executable during compilation, or could be a non-local function that is resolved using the Import Address Table (IAT) \cite{IAT} or in the case of .NET executables is resolved using metadata tables \cite{dotnet}. Statically linked library functions are detected with the help of IDA Pro’s Fast Library Identification and Recognition Technology (FLIRT) \cite{guilfanov1997fast}. After extracting all the API calls inside each function, Mal2GCN gathers the lower case name of every used API and string. If a string has more than 30 characters, we only use the first 30 characters, and if it has less than 4 characters, we throw it away.

\subsection{Feature Vector Embedding}

\begin{figure}[h]
 	\centering
 	\includegraphics[width=0.45\textwidth]{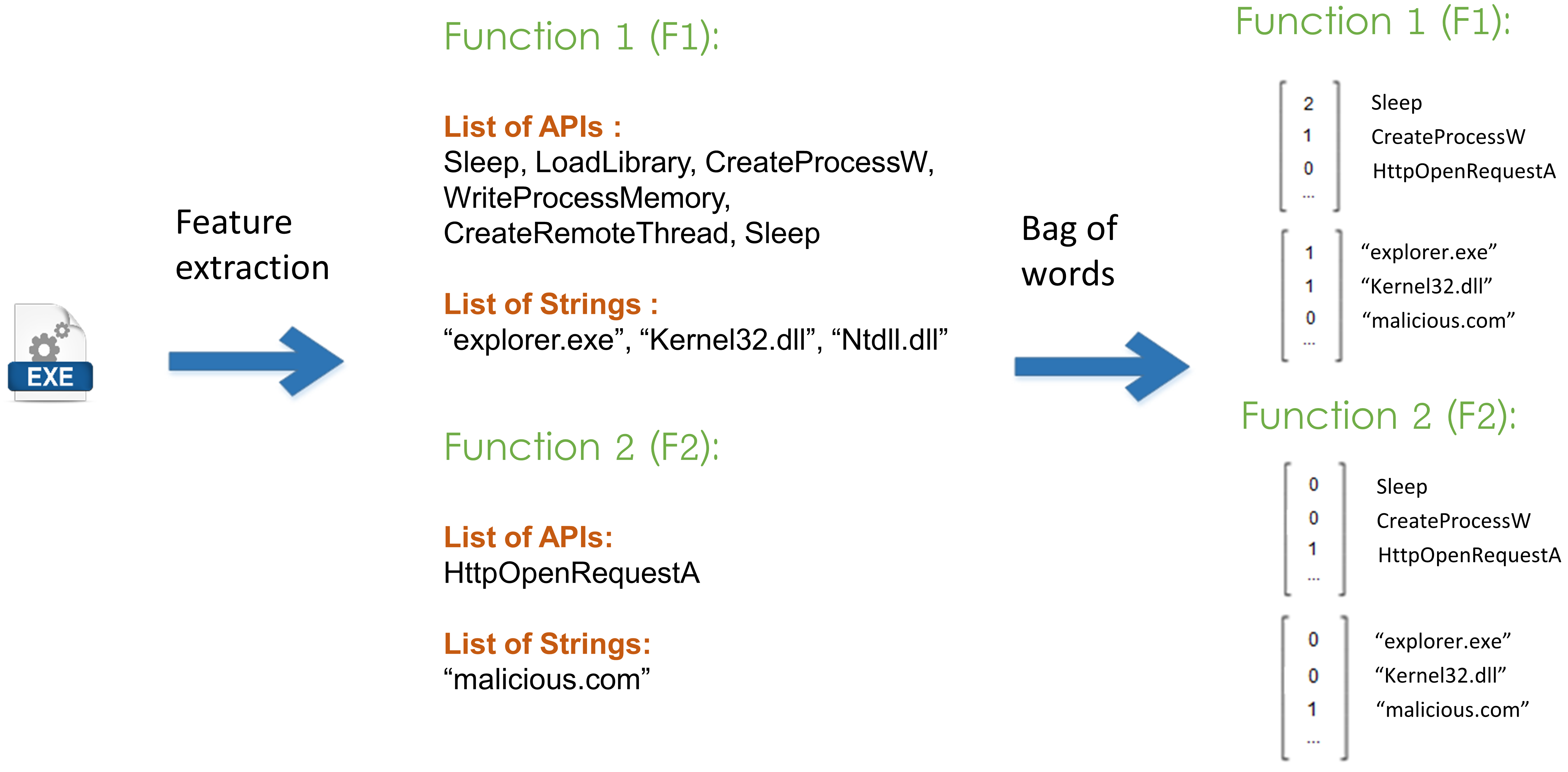}
 	\caption{\label{fig:featureextraction} An example of feature extraction from PE executables, and embedding them into vectors for an executable with two functions.}
 \end{figure}

In order to convert the list of API calls and strings inside each function to a feature vector, Mal2GCN uses the Bag of Words (BoW) approach \cite{manning1999foundations}.
The Bag of Words approach takes a document as input and breaks it into words. These words are also known as tokens, and the process is called tokenization. Unique tokens collected from all processed documents then constitute to form an ordered vocabulary. Finally, a vector with the length of the vocabulary size is created for each document, and every value represents the frequency of a particular token appearing in the corresponding document.\\
For each function, Mal2GCN generates two sentences, one sentence is for the API calls in which every word is the name of one of the API call destinations inside that function, and the other sentence is for all the strings referenced inside that function.
After generating these two sentences for each function, Mal2GCN uses the Bag of Words approach to convert each sentence to a feature vector and finally concatenates these two feature vectors together to generate the initial feature vector for each function. \\
Considering the existence of a very large number of possible APIs and an infinite number of possible strings, we first need to limit the APIs and strings used by Mal2GCN to create the vocabulary. In order to do so, we first gather the most common APIs and strings in the executable files of our training set. Afterward, we use the random forest algorithm to find the top 500 most influential APIs and strings and use them as our vocabulary, so in total, the size of our vocabulary is 1000. Therefore, every function is represented by a vector with the size of 1000. Each entry in the resulting feature vector corresponds to a specific API or a specific string, and its value corresponds to the number of times that API or string has been used in that function. Figure \ref{fig:featureextraction} shows an example of feature extraction and vector embedding.
 
\subsection{Graph Classification}
\begin{figure}[h] 
	\centering
	\includegraphics[width=0.45\textwidth]{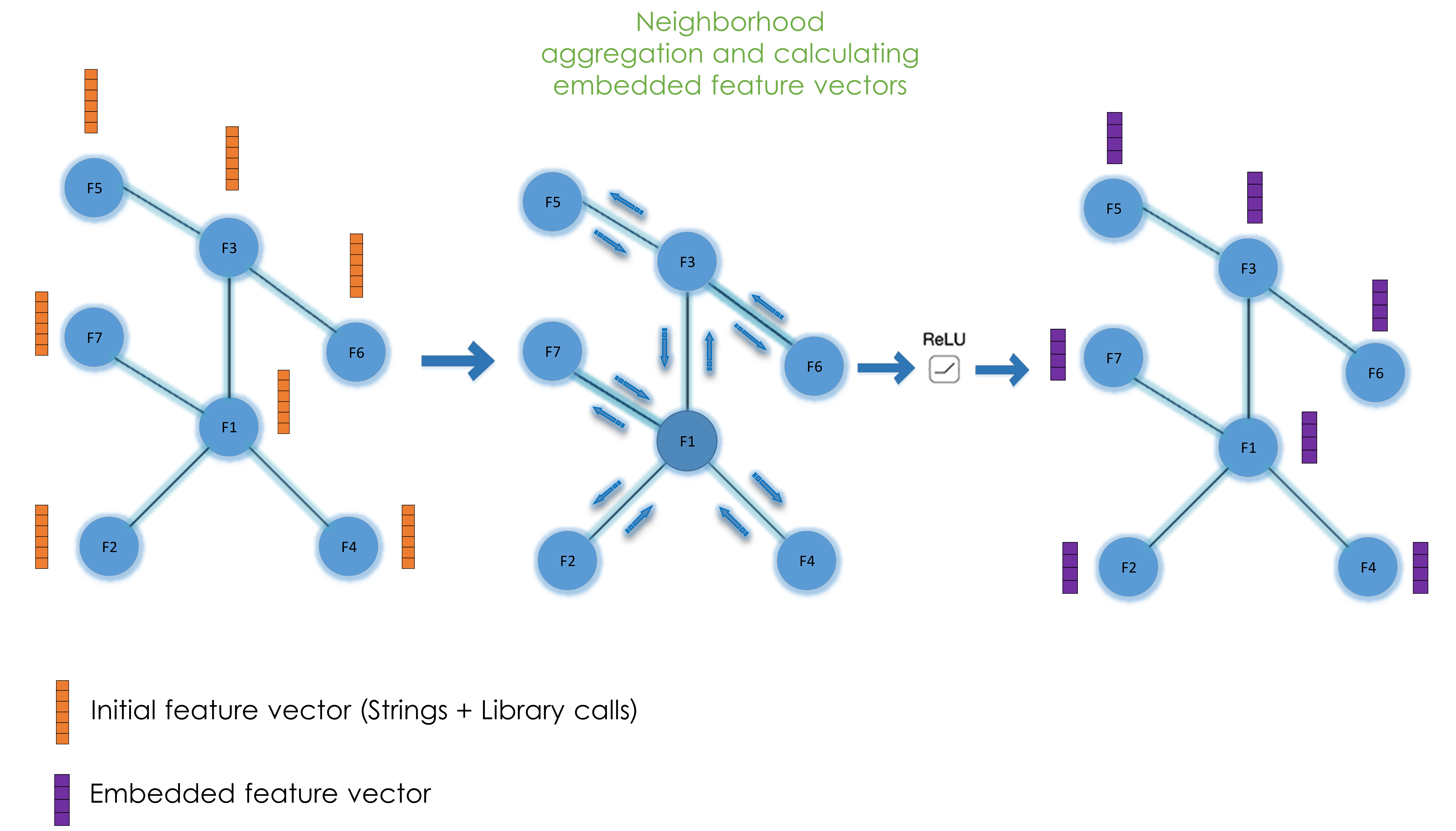}
	\caption{\label{fig:mal2gcn2} Neighborhood aggregation and calculating the node embedding in one layer of GCN.}
\end{figure}

\begin{figure}[h] 
	\centering
	\includegraphics[width=0.45\textwidth]{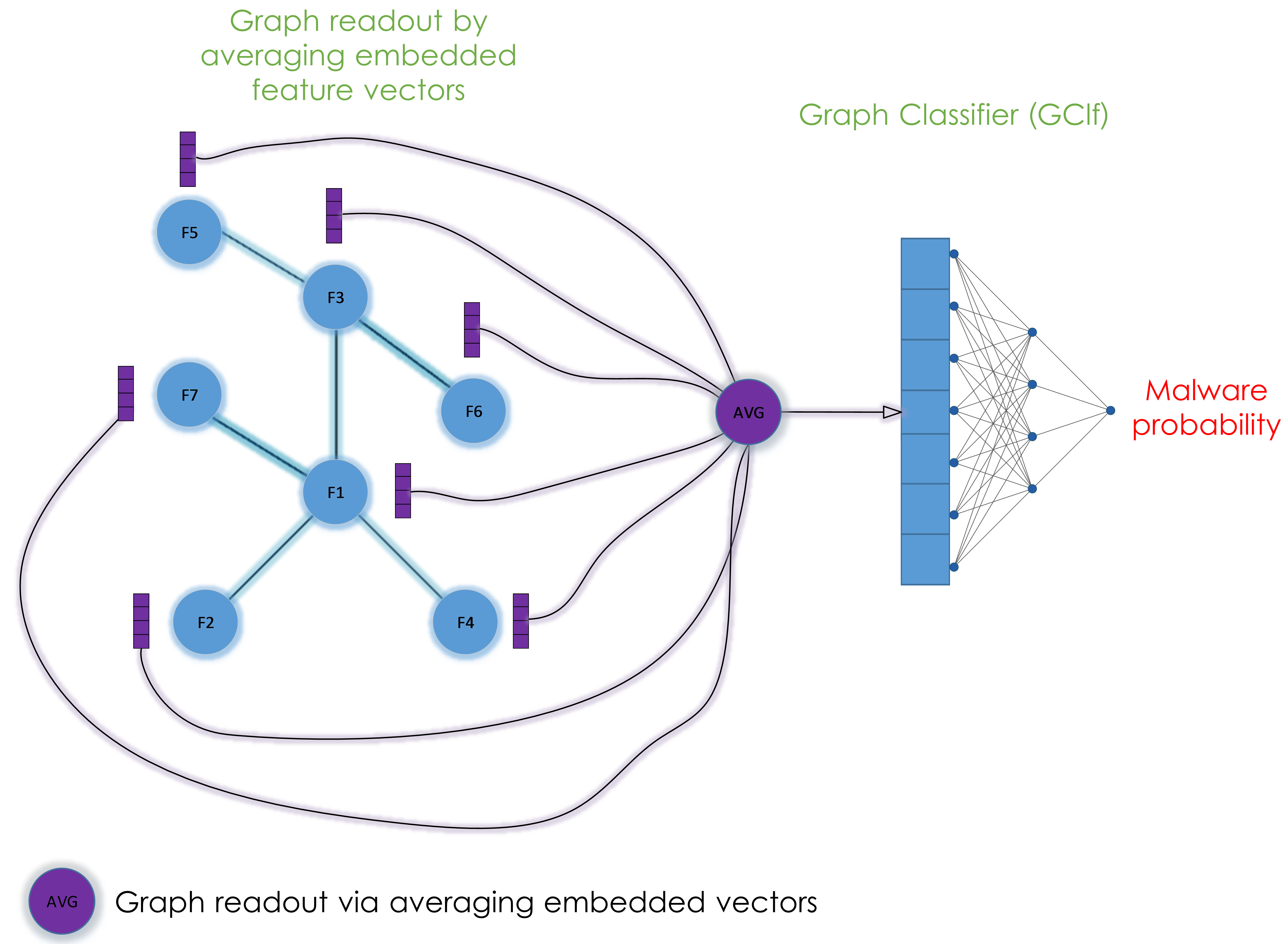}
	\caption{\label{fig:mal2gcn3} Calculating the graph representation vector and classifying the graph using GClf.}
\end{figure}

After calculating the initial feature vector for each function, we construct the Function Call Graph (FCG) for the executable using IDA pro. If any non-library function with a name that includes "main" is found by IDA's automatic analysis, such as "WinMain" or "Main", we mark it as our main function, and if such a function is not found, we mark the function being pointed to by the entry point as our main function. Afterward, if we find any isolated function, which happens when IDA pro cannot find any reference to a function, we connect them to the main function, making sure they can also participate in neighborhood aggregation. The resultant FCG is given to a two-layer GCN, and the output of the GCN is an embedded vector for each node (function). Mal2GCN averages all the embedded feature vectors of nodes and uses the resultant vector as the feature vector of the graph. The graph feature vector is given to a neural network, called \textit{Graph Classifier (GClf)}, to calculate the probability of input being malware. Figures \ref{fig:mal2gcn2} and \ref{fig:mal2gcn3} show an example of classifying a malware using Mal2GCN. \\

\subsection{Mal2GCN Robustness}
\label{sec:malrob}
Evading DL-based malware detectors has a low cost for adversaries, and they can evade these detectors with little effort. Our goal is to increase the cost of evading Mal2GCN by using the FCG representation of executable files and the non-negative training method.
Since FCG representation pertains to malware's functionality and not just the raw byte sequence, a large number of adversarial malware generation approaches, such as the byte appending attacks \cite{kolosnjaji2018adversarial,suciu2019exploring,demetrio2021functionality}, are ineffective in evading Mal2GCN. Appending attacks append bytes at the end or the end of some sections or headers of the PE structure. Since Mal2GCN only takes the referenced strings and APIs in each function as input, such appending attacks do not affect Mal2GCN input and its decision. However, we indicate in section \ref{sec:evaladv} that Mal2GCN is vulnerable to the adversarial source code attack, injecting benign-looking code into malware's source code. In order to make our model robust against such attacks, Mal2GCN uses the non-negative method proposed by Fleshman \textit{et al.} \cite{fleshman2018non} to restrict GCN and GClf only to learn non-negative weights in each layer.  Fleshman \textit{et al.} \cite{fleshman2018non} restrict MalConv weights to non-negative values and propose non-negative MalConv model to defend MalConv model against appending attacks. The authors argue that nothing can be added to an executable file to make it seem more benign to the non-negative MalConv model. In other words, increasing the input of non-negative MalConv can only increase the probability of the input being malware.
Although the authors have expected Non-negative MalConv to be theoretically robust against appending attacks, other studies \cite{ceschin2019shallow,Ebrahimi2020Binary,10.1145/3433210.3453086} demonstrate that Non-negative MalConv is vulnerable to appending attacks. We investigated the vulnerability of non-negative MalConv and found that because of the embedding layer in MalConv, this model cannot be transformed to a monotonically non-decreasing function, even if the embedding layer weights be restricted to non-negative values. 
Since Mal2GCN only includes fully connected and pooling layers, it can be transformed to a monotonically non-decreasing function by restricting its weights to non-negative values. In the following, we indicate that Mal2GCN becomes a monotonically non-decreasing function by restricting its weights to non-negative values.

A neural network is a combination of several layers so that each layer is an affine transformation of inputs followed up by a non-linear function. Neural network $F$ with $m$ layers can be formulated as follows:
\begin{equation}
\begin{split}
&F = F_m \circ F_{m-1} \circ ... \circ F_1 \\
s.t. \quad &\bm{h}^{\ell} = F_{\ell}(h^{\ell-1}) =\phi(W^{\ell} \bm{h}^{\ell-1} + \bm{b}^\ell) 
\end{split}
\end{equation}
where $F_\ell$ is the function of the $\ell^{th}$ layer of neural network, $\bm{h}^\ell$, $W^\ell$, and $\bm{b}^\ell$ are the output, weights, and biases of the $\ell^\text{th}$ layer of neural network, respectively, and $\phi$ is a non-linear function, called activation function. Notably, $\bm{h}^0$ is the input of neural network ($\bm{x}$), and $\bm{h}^m$ is the output of neural network ($\hat{\bm{y}}$). Except for the last layer of GClf, Mal2GCN uses \textit{ReLU} function as the activation function of all layers. Mal2GCN uses \textit{sigmoid} function as the activation function of the last layer of GClf, and thus, the Mal2GCN output $\hat{\bm{y}}$ is in the range [0,1], which indicates the probability of input $\bm{x}$ being malware. 
Since the range of ReLU function is non-negative, and the bag of words method generates non-negative feature vectors, the input of all layers of Mal2GCN is non-negative. In the non-negative training method, the weights of each layer are restricted to non-negative values ($W^\ell \geq 0$).
Regarding the non-negativity of wights and input of each layer, the affine transformation $(W^\ell \bm{h}^{\ell} + \bm{b}^\ell$) is a monotonically non-decreasing function.
Moreover, the ReLU, sigmoid, and pooling functions, such as average, are also monotonically non-decreasing functions.
Therefore, each layer of Mal2GCN with non-negative weights is a monotonically non-decreasing function. Since Mal2GCN with non-negative weights is a combination of several monotonically non-decreasing functions, it is also a monotonically non-decreasing function. Hence, the probability of the input being malware $\hat{\bm{y}}$ is increased by increasing input $\bm{x}$. In other words, an adversary can only increase the probability of the executable being malware by adding benign APIs or strings to the malware source code. 

Non-negative Mal2GCN theoretically guarantees that if the non-adversarial version of malware is classified correctly, injecting or appending benign-looking code inside malware functions cannot fool it. The Non-negativity of Mal2GCN causes the model to only focus on malicious indicators in the function call graph; therefore, adding junk or benign features to the source code, such as benign API calls or strings, can only increase the probability of the input being malware. 
If adversaries want to evade non-negative Mal2GCN, they must change the pattern of malicious behaviors of malware, which greatly increases the cost of attack compared to appending attacks.
To force Mal2GCN to only learn non-negative weights, we convert the negative weights to zeroes after each epoch. This causes the model to focus only on the features that drive the model towards the output of 1 (malware), which are malicious features of the input graph. Through our experiments, we found out that restricting one of the GCN or the GClf is not enough to make Mal2GCN robust, and only when both are restricted to non-negative weights, the model becomes fully robust against benign code injection attacks.\\

\section{Adversarial Source-Code Generation}
\label{sec:Adversarial}
Past research has mostly focused on generating adversarial examples by modifying the compiled executable, but in the real-world, adversaries have access to their own malware source code and they can use various techniques to obfuscate their source code so the compiled executable becomes similar to benign executables and thus avoid detection. Therefore, to evaluate the robustness of malware detection models against real-world adversaries, we need to also evaluate their robustness against source code modification.
In this article, we present a black-box adversarial source code generation approach that injects adversarial code segments into various parts of the source code. The contents of these code segments, such as strings and API calls, are chosen in such a way that makes these injected adversarial code segments very similar to benign source codes. 
All the API calls and strings that are used in these injected code segments change with every injection, and the arguments given as input to these injected API calls changes as well, therefore each generated adversarial malware differs vastly from the previous ones.
Each injected adversarial code segment contains two parts, \textit{Opaque predicates} and \textit{Block content}. 

	\textbf{Opaque predicates}: The content of each adversarial code segment starts with either an \textit{if} or a \textit{while} statement, and the adversarial codes are injected inside the code block of these statements. 
	Before each code block, there will be some calculations, and the result of these calculations will be used in the condition statement of the succeeding \textit{if} or \textit{while}. 
	We adjust the values used in these calculations so that the result of the expression inside the condition statement will always become false. Therefore the content of code segments will never get executed, and thus these injected codes do not change the behavior of the malware. 
	To defeat simple deobfuscation and branch prediction techniques, our adversarial source code generator has the ability to use the returned values from various APIs and environment variables to calculate numbers with a known range and later use these calculated numbers in the condition statements. This means that the deobfuscator needs to properly emulate most of the possible APIs and environment variables, which can be very expensive to predict the result of the condition statement.
	For example, we first use $Environment.OSVersion.Version.Major$ to get the MajorVersion of the underlying MS-windows, which is a number between 0 and 10, and then multiply it by a random number between 1 and 100; therefore, the result will always be between 1 and 1000. We store the resulting value in $variable_{1}$, and also generate another random number that is bigger than 10000 and store it in $variable_{2}$. Finally, inside the condition statement of the succeeding \textit{if} or \textit{while}, we check to see if $variable_{1}$ is bigger than $variable_{2}$, and only execute the code block if it is.
	Therefore this method defeats simple deobfuscation techniques that do not emulate all the possible APIs and environment variables.

	\textbf{Block content}: We inject various benign API calls, string assignments, and random calculations inside the injected code segments. If any of the arguments in any of the injected API calls is a string or an array of strings, we replace it with benign strings, and if it is an integer or float, we replace it with a random number. We use IDA pro \cite{IDA} to collect the list of benign APIs and strings from the benign samples in the training dataset, and the details of which are explained in Section \ref{sec:datasetandsetup}.
	We use the benign APIs and strings inside the injected code segments to make the program's source code similar to benign programs. Doing this will also make the compiled executable similar to benign programs as well, and as we will show in the evaluation section, this technique will cause vulnerable malware detection models to misclassify adversarial malware as benign.

\begin{figure}[h]
	\centering
	\includegraphics[width=0.45\textwidth]{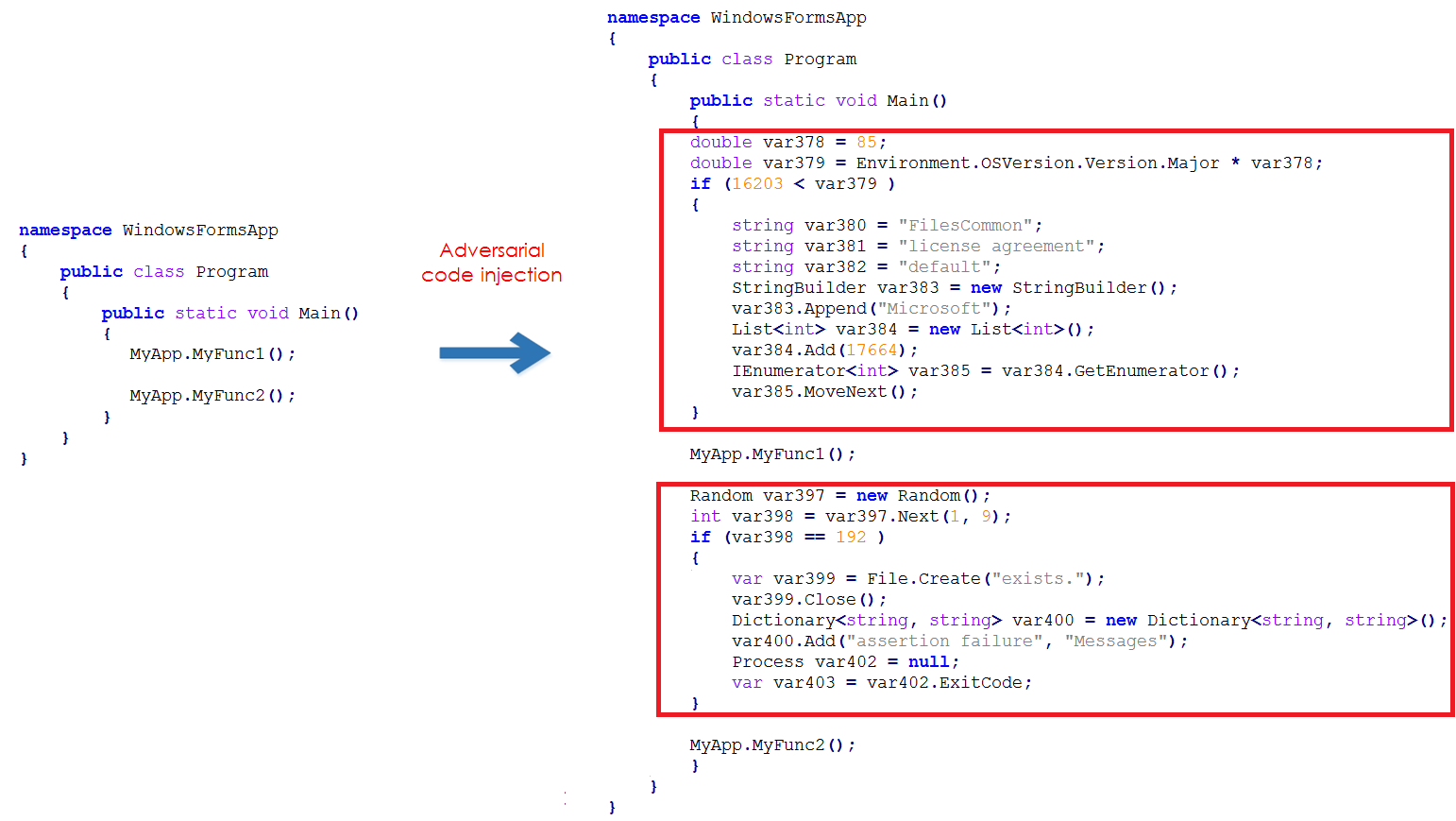}
	\caption{\label{fig:sourcecodeobf} An example of adversarial code injection effect on source code.}
\end{figure}

After generating the adversarial source codes, we compile them to gather the final adversarial executables. Figure \ref{fig:sourcecodeobf} shows an example of injecting two adversarial code segments, and its effect on the source code.

\begin{figure}[h]
	\centering
	\includegraphics[width=0.45\textwidth]{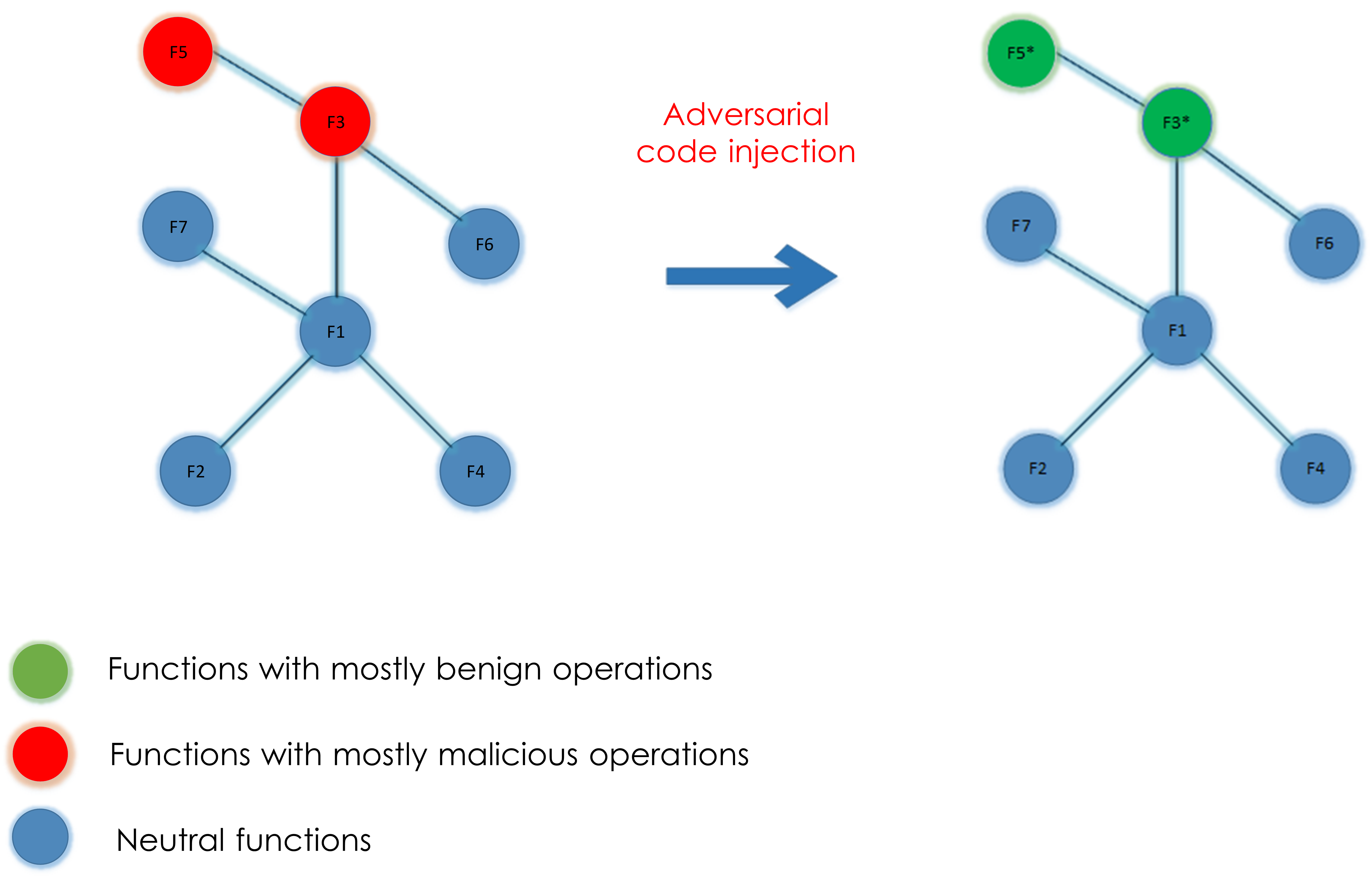}
	\caption{\label{fig:fcgobf} An example of adversarial code injection effect on function call graph (FCG).}
\end{figure}

\begin{figure}[h]
	\centering
	\includegraphics[width=0.45\textwidth]{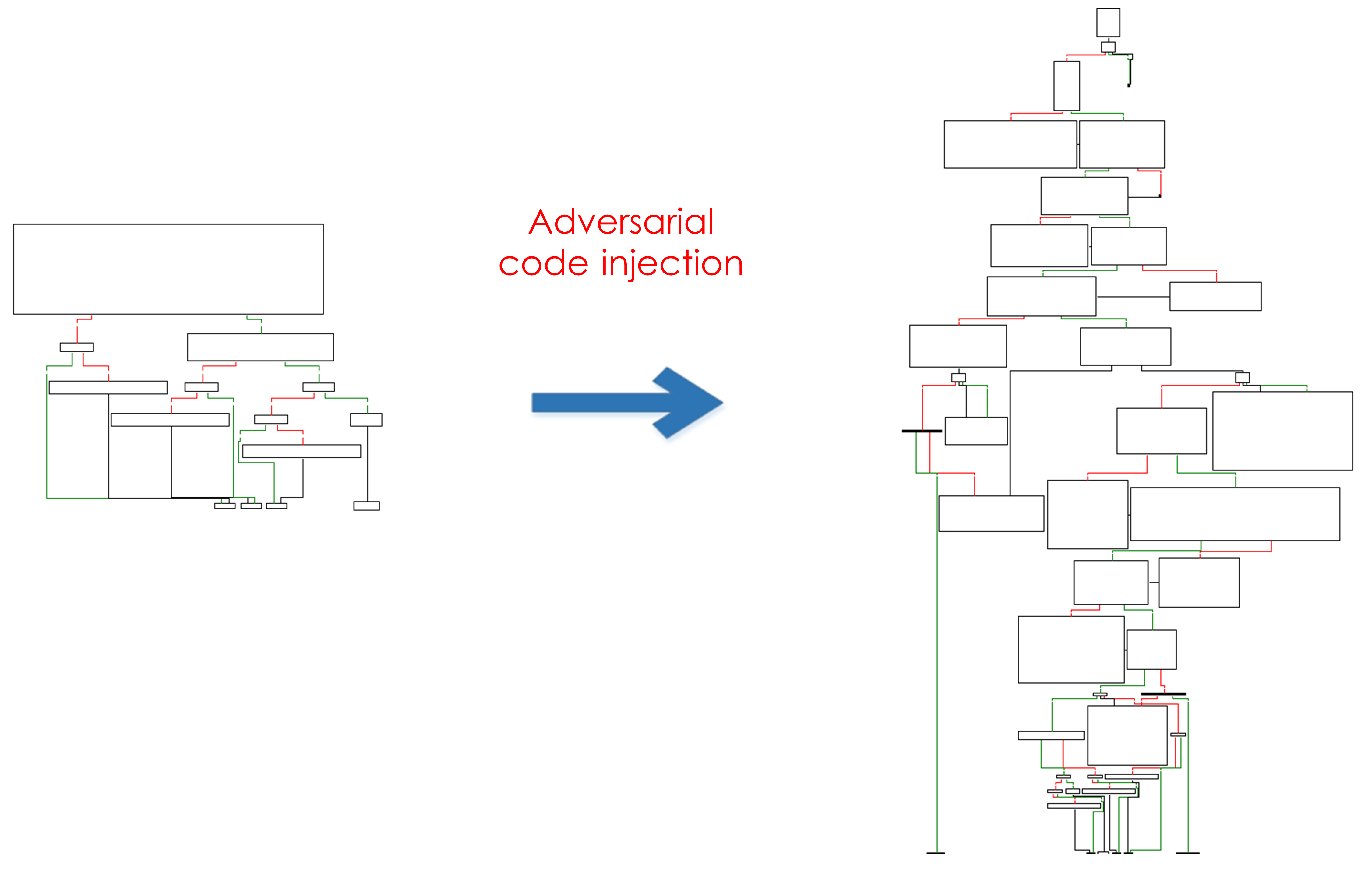}
	\caption{\label{fig:cfgobf} An example of adversarial code injection effect on control flow graph (CFG).}
\end{figure}

Figure \ref{fig:fcgobf} shows an example of adversarial code injection and its effect on function call graph. 
In this figure, red nodes are functions that have mostly malicious operations. We define malicious operations as operations that are mostly used by malware, such as DLL injection or keylogging. Green nodes are functions with mostly benign operations, which are operations that are mostly used by benign programs, such as getting the user input or printing messages, and blue nodes are functions with the same amount of malicious operations as benign operations. The proposed adversarial code generation approach adds adversarial codes into various parts of a malicious function, making it similar to benign functions. In this example, we inject adversarial codes into functions $F1$ and $F2$, causing them to become a function with mostly benign operations.
Figure \ref{fig:cfgobf} shows an example of adversarial code injection effect on the control flow graph of a function. The green lines in this figure show the destination of a basic block if the jump condition is met, red lines are the destination if the jump condition is not met, and black lines are unconditional jumps. Therefore, in the case of injected adversarial codes, the condition is never met and only one of the paths are executed.

\section{Dataset and Setup}
\label{sec:datasetandsetup}
In this section, we first describe how we collect our benign and malware datasets in detail. Afterward, we describe the pre-processing and the training phase of Mal2GCN.

\subsection{Malware Dataset}
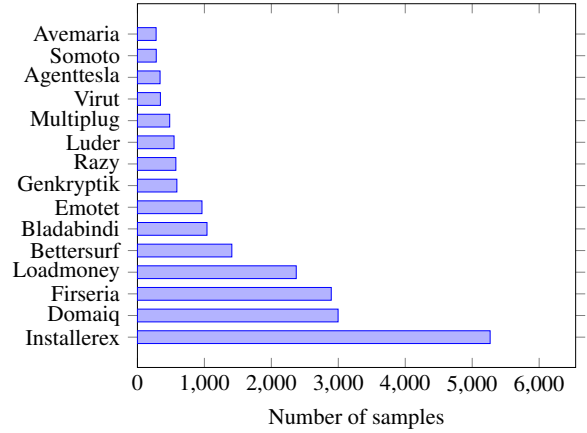
\begin{figure}[h]
	
	\centering
	
\begin{tikzpicture}[scale=0.85]
\centering
\begin{axis}[
xbar,
bar width=0.2cm,
legend style={at={(0.5,-0.15)},
	enlarge x limits={upper,value=0.19},
	xmin=0,xmax=5500,
	anchor=north,legend columns=-1},
xlabel={Malware family},
xlabel={Number of samples},
symbolic y coords={
Installerex,Domaiq,Firseria,Loadmoney,Bettersurf,Bladabindi,Emotet,Genkryptik,Razy,Luder,Multiplug,Virut,Agenttesla,Somoto,Avemaria
},
ytick=data,
nodes near coords align=horizontal,
]
\addplot coordinates {
(5268,Installerex)(2997,Domaiq)(2895,Firseria)(2372,Loadmoney)(1410,Bettersurf)(1039,Bladabindi)(964,Emotet)(589,Genkryptik)(574,Razy)(547,Luder)(482,Multiplug)(343,Virut)(338,Agenttesla)(281,Somoto)(279,Avemaria) 
	
};

\end{axis}

\end{tikzpicture}
	\caption{\label{fig:malwarefamilies} Top 15 most common malware families in our dataset, based on AVclass family names.}
\end{figure}

The malware dataset is obtained through the VirusShare repository \cite{virusshare}. Considering that benign files might also exist in the dataset collected through VirusShare, we also use the VirusTotal service \cite{virustotal} to remove any executable that had less than 20 detections. Afterward, we use AVClass \cite{sebastian2016avclass} to find the family names for the malware in our dataset. Our dataset contains more than 300 different malware families, and the most common malware families are shown in Figure \ref{fig:malwarefamilies}.

\subsection{Benign Dataset}
The common method for creating a benign dataset is to collect the executables of the Windows operating systems. However, only using executables being found on a fresh Windows operating system installation causes the model to become overfitted on specific features that only exist in the system executables of Windows. Hence, it makes the reported results unreliable because the model might only be looking for simple Windows-related features, such as existence of "Microsoft" string in the executable,  to label a file as benign. Also, this method makes the model much easier to evade since adding simple features such as the Windows-related strings can cause the model to be fooled and label a malware as benign.\\
To overcome this issue, we gather benign executables from two different sources:
\begin{itemize}
	\item
	\textbf{Windows system executables:}
	 We gather all the PE executables in the fresh installation of different versions of Windows, from Windows XP to Windows 10, including 32-bit and 64-bit executables. Considering there are various types of PE files, we only gathered PE files that have the \textit{exe}, \textit{dll}, or \textit{sys} extensions.
	
	\item
	\textbf{Crawled benign executables:}
	 We write a crawler to collect executable files from various software downloading websites such as freewarefiles \cite{freewarefiles}. One important thing to consider regarding the downloaded executable from these websites is that most of them are installers and not the real executable of the application. For example, most developers will use installers such as Inno Setup \cite{Inno} to compress and pack various PE files of their application into one installer executable. Therefore, only using these collected files without unpacking them will cause the model to get overfitted to simple features in specific installers. To overcome this issue, we use UniExtractor \cite{uniextractor} to unpack the executables that exist in the crawled files and gather the real executables that were packed inside the installers.
After the unpacking phase and gathering all the benign executables, we use the VirusTotal service to remove any executable with more than two detections.
	
\end{itemize}

\begin{figure}
	\begin{subfigure}
		
		\centering{
			\begin{tikzpicture}
			\begin{groupplot}[group style={group size= 1 by 1,vertical sep=2.5cm},height=5cm,width=8cm,tick label style = {font = {\fontsize{8 pt}{12 pt}\selectfont}},]
			\nextgroupplot[ymin=0,xmax=1000, ymax=9999,enlargelimits=true,enlargelimits=0.00001,ylabel=\footnotesize{\# of executable},xlabel=\footnotesize{Number of functions},ytick={0,2000,4000,6000,8000,9999},yticklabels={0,2K,4K,6K,8K,10K}]
			
			\addplot
			[ybar interval,mark=no,fill=purple] 
			coordinates
			{
				( 0 , 2901 )( 20 , 3269 )( 40 , 1815 )( 60 , 2601 )( 80 , 1753 )( 100 , 1027 )( 120 , 966 )( 140 , 846 )( 160 , 733 )( 180 , 634 )( 200 , 617 )( 220 , 558 )( 240 , 582 )( 260 , 1457 )( 280 , 451 )( 300 , 418 )( 320 , 400 )( 340 , 323 )( 360 , 657 )( 380 , 322 )( 400 , 311 )( 420 , 380 )( 440 , 327 )( 460 , 298 )( 480 , 370 )( 500 , 254 )( 520 , 217 )( 540 , 220 )( 560 , 233 )( 580 , 219 )( 600 , 176 )( 620 , 223 )( 640 , 156 )( 660 , 150 )( 680 , 159 )( 700 , 134 )( 720 , 153 )( 740 , 135 )( 760 , 132 )( 780 , 122 )( 800 , 86 )( 820 , 88 )( 840 , 73 )( 860 , 52 )( 880 , 48 )( 900 , 33 )( 920 , 19 )( 940 , 29 )( 960 , 28 )( 980 , 27 )
			} 
			\closedcycle;

			\end{groupplot}
			\end{tikzpicture}
		}
		\vspace{-1.2\baselineskip}
		\caption{ Histogram of the number of functions in the benign dataset.}
		\label{fig:histbenign}
	\end{subfigure}
	
	\vspace{0.5cm}
	
	\begin{subfigure}
		
		\centering{
			\begin{tikzpicture}
			\begin{groupplot}[group style={group size= 1 by 1,vertical sep=1.5cm},height=5cm,width=8cm,tick label style = {font = {\fontsize{8 pt}{12 pt}\selectfont}},]
			\nextgroupplot[ymin=0, ymax=9999,xmax=1000,enlargelimits=true,enlargelimits=0.00001,ylabel=\footnotesize{\# of executable},xlabel= \footnotesize{Number of functions},ytick={0,2000,4000,6000,8000,9999},yticklabels={0,2K,4K,6K,8K,10K}]
			
			\addplot
			[ybar interval,mark=no,fill=purple] 
			coordinates
			{
				( 0 , 3162 )( 20 , 9728 )( 40 , 1605 )( 60 , 3761 )( 80 , 896 )( 100 , 723 )( 120 , 636 )( 140 , 489 )( 160 , 537 )( 180 , 790 )( 200 , 331 )( 220 , 3334 )( 240 , 285 )( 260 , 140 )( 280 , 150 )( 300 , 107 )( 320 , 152 )( 340 , 198 )( 360 , 269 )( 380 , 61 )( 400 , 202 )( 420 , 83 )( 440 , 2283 )( 460 , 55 )( 480 , 113 )( 500 , 74 )( 520 , 38 )( 540 , 314 )( 560 , 43 )( 580 , 40 )( 600 , 46 )( 620 , 72 )( 640 , 23 )( 660 , 29 )( 680 , 23 )( 700 , 13 )( 720 , 26 )( 740 , 12 )( 760 , 41 )( 780 , 10 )( 800 , 45 )( 820 , 3 )( 840 , 4 )( 860 , 16 )( 880 , 3 )( 900 , 5 )( 920 , 1 )( 940 , 4 )( 960 , 0 )( 980 , 0 )
			} 
			\closedcycle;

			\end{groupplot}
			\end{tikzpicture}
		}
		\vspace{-1.2\baselineskip}
		\caption{ Histogram of the number of functions in the malware dataset.}
		\label{fig:histmalware}
	\end{subfigure}
\end{figure}

\subsection{Preprocessing and Data Gathering}	
The final dataset contains 58157 PE files, from which 30975 PE files are malware, and 27182 PE files are benign. We develop an IDA Pro script for collecting various types of information from the executable files in our dataset. To use the script efficiently, we also develop a python program with multi-processing capabilities to automatically execute our IDA Pro script for a large number of executables. Using our IDA pro script, we collect three types of information from the executable files: 

\begin{enumerate}
	\item
List of functions being called by each function
\item
API calls used in each function
\item 
Strings referenced in each function
\end{enumerate}
The list of local function calls in each function is used to construct the function call graph, and the API calls and referenced strings are used to calculate the feature vector for each function. The histograms of the number of functions for benign and malware executables are presented in Figure \ref{fig:histbenign} and Figure \ref{fig:histmalware}, respectively.

\subsection{Generating Adversarial Malware}

We first need to select a malware whose source code is available to generate adversarial malware using the adversarial source code generating method. We use the source code of \textit{Lime Crypter} \cite{lime} to generate adversarial malware. Crypters are a rising threat to anti-malware products and are used by cybercriminals to bypass traditional signature detection methods \cite{yan2008revealing,holt2016cybercrime,balci2016art}, and many known malware families use them to bypass detection methods \cite{Raccoon1}. Considering that some of the advanced crypters also use similar methods to the proposed adversarial source code generation method \cite{beek2017mcafee}, we specifically selected a crytper for generating malware to evaluate the models against this real-world adversary.
We write a python script that injects adversarial codes into various parts of Lime Crypter source code to generate adversarial source codes. Afterward, we compile all the generated adversarial malware source codes with the help of batch scripting and generate the final adversarial malware. We also turn off the optimization while compiling the adversarial source codes to make sure non of the injected code segments gets removed because of the compiler's optimization.

To gather APIs and strings being injected inside the adversarial code segments, we collect the most common APIs and strings in the benign Windows system executables in our dataset. We chose to only use Windows system executable files, because any attacker in the real world also has access to these files, considering that the attacker only needs to install different versions of Windows operating system in order to gather these executable files, just as we did. Considering that Lime Crypter source code is written in C\#, we first use Detect It Easy \cite{die} to find .NET based executables, and then use our IDA pro script to gather the most common APIs among them, and then use the strings tool to find the ascii and unicode strings. Finally, we collect two lists, the list of benign strings and the list of benign APIs, and use them to generate the content of the adversarial code segments.  Each generated adversarial malware can have from 10\% to 500\% overhead in terms of lines that are added to its source code, and the location of injected adversarial codes changes in each sample.
\section{Evaluation}
\label{sec:Evaluation}
We use Deep Graph Library (DGL) \cite{wang2019deep} on top of the PyTorch platform to train Mal2GCN. PE dataset is split into train, validation, and test sets, and they contain 38977, 9590, and 9590 PE files, respectively.
In order to generate the DGL graphs for each program, we use the function call lists being gathered using IDA pro and iterate through all of these lists to add edges based on local function calls. 
We use the Adam optimizer with the batch size of 32 and early stopping with the patience of 3 to train Mal2GCN.
We fine-tune the other hyperparameters of Mal2GCN using the validation set. The list of hyperparameters and their chosen values is shown in Table \ref{table:hyper}.
The default values proposed in \cite{raff2018malware,fleshman2018non} are used to train the MalConv and the non-negative MalConv. All models are trained in a maximum of 100 epochs.
All experiments are done on a machine with an Intel Core i7-6700k CPU, Geforce GTX 980Ti GPU, and 32 GB RAM. We first evaluate the performance of MalConv and Mal2GCN in various settings on the test set of our dataset and then evaluate the robustness of those models against adversarial malware.

\begin{table}[]
	\caption{Hyperparameters Selection for Mal2GCN.}
	\label{table:hyper}
	\renewcommand{\arraystretch}{1.7}
	\resizebox{\linewidth}{!}{
		\begin{tabular}{lcc}
			\hline
			Hyperparameter                      & Search Range             & Best Value \\ \hline
			\# of GCN layers                    & {[}1, 2, 3{]}            & 2          \\
			\# of GClf hidden layers  & {[}0, 1, 2, 3{]}         & 1          \\
			GCN layer 1 size                    & {[}10, 250, 500, 1000{]} & 500        \\
			GCN layer 2 size                    & {[}10, 250, 500, 1000{]} & 250        \\
			GClf hidden layer size & {[}32, 64, 128, 256{]}   & 64         \\
			Graph readout type                  & {[}avg, sum, max{]}      & avg        \\
			Learning rate                       & {[}0.008, 0.08, 0.8{]}   & 0.008      \\ \hline
		\end{tabular}
	}
\end{table}

\subsection{MalConv and Mal2GCN Performance}

\begin{table}[]
	\caption{Comparision of Accuracy, Precision, Recall, and F1 Score of Mal2GCN and MalConv in Various Settings.}
	\label{table:models}
	\centering
	\renewcommand{\arraystretch}{1.4}
	\resizebox{\linewidth}{!}{
		\begin{tabular}{lcccc}
			\hline
			Model      & \multicolumn{1}{l}{Accuracy(\%)} & \multicolumn{1}{l}{Precision(\%)} & \multicolumn{1}{l}{Recall(\%)} & \multicolumn{1}{l}{F1 Score(\%)}  \\ \hline
			MalConv    & 94.98                            & 98.17                             & 93.14                          & 95.59                                           \\
			MalConv-AT & 95.72                            & 98.27                             & 94.34                          & 96.26                                         \\
			MalConv+   & 90.50                            & \textbf{99.26}                             & 84.37                          & 91.21                                   \\
			Mal2GCN    & 97.44                            & 98.44                             & 97.16                          & 97.80                                               \\
			Mal2GCN-AT & \textbf{98.15}                   & 99.06                    & \textbf{97.76}                 & \textbf{98.41}                                              \\
			Mal2GCN+   & 96.41                            & 98.68                             & 95.12                          & 96.87                                            \\ \hline
		\end{tabular}
	}
\end{table}

Table \ref{table:models} shows the performance of MalConv and Mal2GCN in terms of accuracy, precision, recall, and F1 score in various settings. When a model is trained in a non-negative fashion, we denote it by appending “+” to its name, and when a model is adversarially trained, we denote it by appending "-AT" to its name.
The models in Table \ref{table:models} are explained in the following: 

\begin{itemize}
	
	\item
	\textbf{Mal2GCN}: The Mal2GCN model is normally trained, and the model is free to learn any weight.
	\item
	\textbf{Mal2GCN-AT:} Besides natural samples, we also give 2000 adversarial malware generated using our approach to the model during training.
	\item 
	\textbf{Mal2GCN+:} We apply the non-negative weight restriction to the GCN and the GClf layers during training.
	\item 
	\textbf{MalConv:}  The MalConv model is trained as explained in \cite{raff2018malware}, and the model is free to learn any weight.
	\item 
	\textbf{MalConv-AT:} The same aforementioned 2000 adversarial malware is used to adversarially train the MalConv model.
	\item 
	\textbf{MalConv+:} The non-negative weight restriction is applied to all the layers of the MalConv model during training, as explained in \cite{fleshman2018non}.
\end{itemize}

As shown in Table \ref{table:models}, Mal2GCN outperforms MalConv in all of the settings. 
The performance hit of restricting the model to non-negative weights is vastly reduced in Mal2GCN, and using this restriction will only cause the reduction of 1\% in the accuracy, therefore making it more practical to use. Figure \ref{fig:ROC} shows the ROC curves and AUC of MalConv and Mal2GCN in various settings. Table \ref{table:runtime} shows the runtime of every step in Mal2GCN. Note that generating the graphs using IDA pro is optional, and the input graphs can be generated using any tool or framework. In a real-world scenario, such as in a malware detection engine, these graphs can be generated using highly optimized algorithms, which are much faster than running an IDA pro script. Nevertheless, as runtime has not been our main concern, we used IDA pro to generate the graphs.
  
    \input{content/roc.tex}
  
  \begin{table}[]
  	\caption{Average of runtime in different steps of the classification.}
  	\label{table:runtime}
  	\renewcommand{\arraystretch}{1.7}
  	\resizebox{\linewidth}{!}{
  		\begin{tabular}{lc}
  			\hline
  			\multicolumn{1}{c}{Step}                                                                     & Avg Runtime        \\ \hline
  			\begin{tabular}[c]{@{}l@{}}Call graph and feature extraction  using IDA pro\end{tabular}   & 0.28 sec. per app  \\
  			Generating DGL graphs                                                                        & 0.025 sec. per app \\
  			\begin{tabular}[c]{@{}l@{}}Classifying using Mal2GCN with DGL graphs as input\end{tabular} & 0.014 sec. per app \\ \hline
  		\end{tabular}
  	}
  \end{table}

  \subsection{Robustness  Against Adversarial Malware}
  \label{sec:evaladv}
\begin{figure}
	\centering
\begin{tikzpicture}
\begin{axis}[
xmin = -20,
xmax = 520,
height=7cm,
width=8cm,
xlabel = \scriptsize{Adversarial Source Code Overhead (\%)},
ylabel = \scriptsize{Adversarial Malware Success Rate (\%)},
title = \small{},
grid=major,
,legend style={at={(0.9,0.1)},anchor=south east},
xtick= {0,50,100,200,300,400,500}
]

\addplot[blue, mark = *,mark size = 1] coordinates {
	(0, 2)
	(5, 12)
	(10, 34)
	(20, 49)
	(30, 94)
	(40, 99)
	(50, 100)
	(100, 100)
	(150, 100)
	(200, 100)
	(400, 100)
	(500,100 )
};
\addlegendentry{MalConv}

\addplot[red, mark =x,mark size = 1] coordinates {
	(0, 0)
	(5, 0)
	(10, 0)
	(20, 0)
	(30, 0)
	(40, 2)
	(50, 12)
	(100, 31)
	(150, 76)
	(200, 84)
	(400, 95)
	(500,98 )
};
\addlegendentry{Mal2GCN}
\end{axis}
\end{tikzpicture}
\caption{Adversarial malware success rate with different overheads. Overheads are calculated based on the number of adversarial lines that are added to the source code.}
\label{fig:advsuccessrate}
\end{figure}
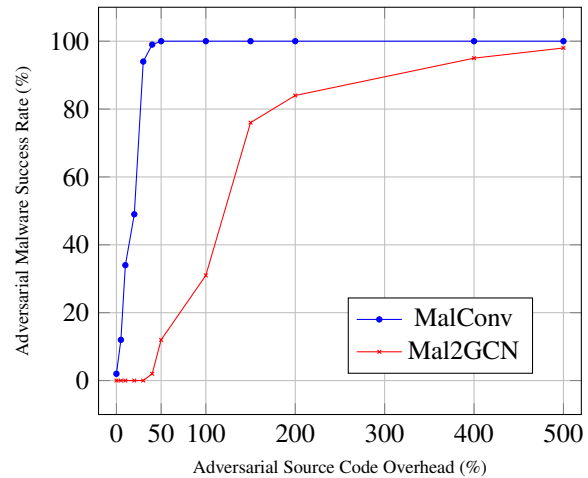

  We generate 2000 adversarial malware using the proposed adversarial code injection approach and evaluate the robustness of models against them. Table \ref{tab:robacc} shows the accuracy of MalConv and Mal2GCN on adversarial malware, called robust accuracy. The result demonstrates that Mal2GCN is more robust than MalConv in all settings.
 Mal2GCN+ has the highest robust accuracy and is able to detect 100\% of the generated adversarial malware.
 It is also shown that adversarial training is not enough to make the models robust against adversarial malware, and using non-negative weights is a much better defense against complex attacks in terms of robustness. 
 We think the main reason for the lack of robustness in adversarially trained malware detection models is the high number of ways an adversary can perturb the malware. In the image classification domain, it is supposed that the perturbation size is bounded by a distance metric, such as $L_P$-norms. However, there is no restriction on the size, content, and location of perturbation in the malware detection domain, and an adversary can generate several adversarial malware being vastly different from each other.
 As shown in the results, the generated adversarial malware also evade the MalConv+ model with a success rate of 99.92\%, having a higher evasion success rate than attacking MalConv. We conjecture that this is because of the fact that the proposed adversarial malware generation approach injects codes into various parts of the source code, causing the compiled executable and its corresponding byte sequence to be vastly different from the non-adversarial executables; therefore, many of the sequences of bytes that MalConv looks for as an indicator of maliciousness no longer exist, and using non-negative weights in this scenario will only cause the reduction of performance. To evaluate the impact of non-negativity on the robustness of Mal2GCN, we consider two other versions of Mal2GCN that are explained in the following:
  
  \begin{itemize}
  \item
  \textbf{Mal2GCN-GClf+}: This model enforces the non-negative weights only on the GClf layers.
  \item
  \textbf{Mal2GCN-GCN+}: This model enforces the non-negative weights only on the GCN layers.
	\end{itemize}
  As shown in table \ref{tab:mal2gcnversions}, enforcing non-negative weights at the GCN layers causes the model to become very robust against the generated adversarial malware, but still very few of them can bypass the model. If the non-negative weights are enforced at GClf layers, the model becomes even less robust than the default Mal2GCN model, which is interesting and shows that GCN layers have more impact on the robustness of Mal2GCN.\\
We also measured the amount of overhead that is required for each default model to get bypassed, and the results are shown in Figure \ref{fig:advsuccessrate}. The overhead is based on the number of adversarial lines added to the malware source code in this figure.
As seen, MalConv is fully bypassed with less than 50\% overhead, but to fully bypass the default Mal2GCN, more than 500\% overhead is required, which shows that even without the non-negative constraints, Mal2GCN is still hard to bypass for adversaries. The results of Table \ref{tab:robacc} and Figure \ref{fig:advsuccessrate} demonstrate that FCG representation is more robust than raw byte sequence representation of executable files for malware detection.

   \textbf{Robustness against other attacks}: Many other adversarial malware attacks have been proposed in recent years. Most of them are based on the idea of appending attacks \cite{kolosnjaji2018adversarial,suciu2019exploring,demetrio2021functionality}. As mentioned in Section \ref{sec:malrob}, such appending attacks do not affect Mal2GCN because they do not change the FCG of executable files.
   Note that Mal2GCN detects a call instruction as an API call when it properly references a function inside the current PE's Import Address Table (IAT), or in the metadata tables in case of .NET executables, or when the destination is the start of a statically linked library function which was detected by FLIRT. Moreover, these call instructions need to be inside a local function located by IDA pro.
   Therefore, appending bytes at the end of executable sections will not change the FCG of executable files and thus the Mal2GCN decision. There are also other attacks that use reinforcement learning and Generative Adversarial Networks (GAN) to modify the PE structure, but as these attacks do not modify the source code and the referenced strings and APIs inside each function \cite{anderson2018learning,pesidious,hu2017generating}, they do not modify the FCG of executable files either. The only modification in these attacks that could affect Mal2GCN is packing it with UPX, which can be easily defeated by unpacking the sample before giving it to Mal2GCN. 
  
 \begin{table}[]
 	\caption{The robust accuracy of Mal2GCN and MalConv in various settings.}
 	\label{tab:robacc}
 			\renewcommand{\arraystretch}{1.7}
 	\resizebox{\linewidth}{!}{
 	\begin{tabular}{cccc}
 		& \multicolumn{3}{c}{Robust Accuracy (\%)}                                 \\ \cline{2-4} 
 		Models  & Normally Trained & Adversarially Trained (AT) & Non-Negative Weights (+)\\ \hline
 		MalConv & 7.15             & 35.65                      & 0.08                     \\
 		Mal2GCN  & 40.05            & 75.20                      & 100                      \\ \hline
 	\end{tabular}
}
 \end{table}

\begin{table}[]
	\caption{Comparison of robust accuracy when enforcing non-negative weights at different parts of the Mal2GCN model}
	\label{tab:mal2gcnversions}
	\resizebox{\linewidth}{!}{
	\begin{tabular}{cccc}
		\multicolumn{4}{c}{Robust Accuracy (\%)}        \\ \hline
		Mal2GCN & Mal2GCN-GClf+ & Mal2GCN-GCN+ & Mal2GCN+ \\ \hline
		40.05   & 8.55        & 98.85        & 100      \\ \hline
	\end{tabular}
}
\end{table}

\section{Discussion and Limitations}
\label{sec:Discussion}
Although we showed that Mal2GCN+ could resist junk/benign code injection attacks, just like any other static-based model, this model is vulnerable against attacks that target static malware detector limitations, such as its weakness against packed executables \cite{swinnen2014one}. Adversaries can also use other methods to bypass static malware detectors, such as dynamically loading libraries, calling their functions on runtime, finding the library function addressed by parsing the Thread Environment Block (TEB) \cite{snow2012automatic}, and encrypting strings and decrypting them on runtime. As a workaround, we could run the program in a sandbox to capture the function call graph and APIs during runtime, but running a program inside a sandbox is not always possible, especially when analyzing a large number of files. The solution to this limitation is using emulators to emulate the program for a maximum number of instructions, and therefore overcome the limitation of static methods and also be able to analyze a large number of files \cite{green2015detecting,golshan2016systems,kang2009emulating}.
\section{Conclusion and Future Work}
\label{sec:Conclusion}
In this paper, we first presented Mal2GCN, a robust and accurate malware detection model that uses Graph Convolutional Network (GCN) with non-negative weights. Mal2GCN uses Function Call Graph (FCG) of executable files. We demonstrated that FCG representation of executable files is more robust than raw byte sequence representation. Since FCG representation does not rely on the raw byte sequence of executable files and is related to the true functionality of the malware, numerous adversarial malware generating methods, such as appending attacks, are ineffective in evading Mal2GCN. We also used the non-negative training method to increase the robustness of Mal2GCN against our proposed attack that injects adversarial code into the various part of the malware source code without altering the true functionality of the malware.
In future works, we will present Emu2GCN, which will use emulation to generate the FCG of executable files. This approach will eliminate many shortcomings of static malware detection methods, including their vulnerability against custom packers and API obfuscations. 



\normalsize
\bibliography{mybib}

\begin{thebibliography}{73}
\providecommand{\natexlab}[1]{#1}
\providecommand{\url}[1]{\texttt{#1}}
\expandafter\ifx\csname urlstyle\endcsname\relax
  \providecommand{\doi}[1]{doi: #1}\else
  \providecommand{\doi}{doi: \begingroup \urlstyle{rm}\Url}\fi

\bibitem[Goodfellow et~al.(2014)Goodfellow, Shlens, and
  Szegedy]{goodfellow2014explaining}
Ian~J Goodfellow, Jonathon Shlens, and Christian Szegedy.
\newblock Explaining and harnessing adversarial examples.
\newblock \emph{arXiv preprint arXiv:1412.6572}, 2014.

\bibitem[Akhtar and Mian(2018)]{akhtar2018threat}
Naveed Akhtar and Ajmal Mian.
\newblock Threat of adversarial attacks on deep learning in computer vision: A
  survey.
\newblock \emph{IEEE Access}, 6:\penalty0 14410--14430, 2018.

\bibitem[Brendel et~al.(2017)Brendel, Rauber, and Bethge]{brendel2017decision}
Wieland Brendel, Jonas Rauber, and Matthias Bethge.
\newblock Decision-based adversarial attacks: Reliable attacks against
  black-box machine learning models.
\newblock \emph{arXiv preprint arXiv:1712.04248}, 2017.

\bibitem[Ilyas et~al.(2018)Ilyas, Engstrom, Athalye, and Lin]{ilyas2018black}
Andrew Ilyas, Logan Engstrom, Anish Athalye, and Jessy Lin.
\newblock Black-box adversarial attacks with limited queries and information.
\newblock \emph{arXiv preprint arXiv:1804.08598}, 2018.

\bibitem[Papernot et~al.(2016)Papernot, McDaniel, Jha, Fredrikson, Celik, and
  Swami]{papernot2016limitations}
Nicolas Papernot, Patrick McDaniel, Somesh Jha, Matt Fredrikson, Z~Berkay
  Celik, and Ananthram Swami.
\newblock The limitations of deep learning in adversarial settings.
\newblock In \emph{2016 IEEE European symposium on security and privacy
  (EuroSP)}, pages 372--387. IEEE, 2016.

\bibitem[Suciu et~al.(2019)Suciu, Coull, and Johns]{suciu2019exploring}
Octavian Suciu, Scott~E Coull, and Jeffrey Johns.
\newblock Exploring adversarial examples in malware detection.
\newblock In \emph{2019 IEEE Security and Privacy Workshops (SPW)}, pages
  8--14. IEEE, 2019.

\bibitem[Hu and Tan(2017)]{hu2017generating}
Weiwei Hu and Ying Tan.
\newblock Generating adversarial malware examples for black-box attacks based
  on gan.
\newblock \emph{arXiv preprint arXiv:1702.05983}, 2017.

\bibitem[Anderson et~al.(2018)Anderson, Kharkar, and
  Filar]{anderson2018learning}
Hyrum~S Anderson, Anant Kharkar, and EndGame~Inc Filar.
\newblock Learning to evade static pe machine learning malware models via
  reinforcement learning.
\newblock \emph{arXiv preprint arXiv:1801.08917}, 2018.

\bibitem[Kolosnjaji et~al.(2018)Kolosnjaji, Demontis, Biggio, Maiorca,
  Giacinto, Eckert, and Roli]{kolosnjaji2018adversarial}
Bojan Kolosnjaji, Ambra Demontis, Battista Biggio, Davide Maiorca, Giorgio
  Giacinto, Claudia Eckert, and Fabio Roli.
\newblock Adversarial malware binaries: Evading deep learning for malware
  detection in executables.
\newblock In \emph{2018 26th European signal processing conference (EUSIPCO)},
  pages 533--537. IEEE, 2018.

\bibitem[Demetrio et~al.(2021{\natexlab{a}})Demetrio, Coull, Biggio, Lagorio,
  Armando, and Roli]{demetrio2021adversarial}
Luca Demetrio, Scott~E Coull, Battista Biggio, Giovanni Lagorio, Alessandro
  Armando, and Fabio Roli.
\newblock Adversarial exemples: a survey and experimental evaluation of
  practical attacks on machine learning for windows malware detection.
\newblock \emph{ACM Transactions on Privacy and Security (TOPS)}, 24\penalty0
  (4):\penalty0 1--31, 2021{\natexlab{a}}.

\bibitem[Maiorca et~al.(2020)Maiorca, Demontis, Biggio, Roli, and
  Giacinto]{maiorca2020adversarial}
Davide Maiorca, Ambra Demontis, Battista Biggio, Fabio Roli, and Giorgio
  Giacinto.
\newblock Adversarial detection of flash malware: Limitations and open issues.
\newblock \emph{Computers and Security}, 96:\penalty0 101901, 2020.

\bibitem[Li and Li(2021)]{li2021irl}
Xintong Li and Qi~Li.
\newblock An irl-based malware adversarial generation method to evade
  anti-malware engines.
\newblock \emph{Computers and Security}, 104:\penalty0 102118, 2021.

\bibitem[Al-Dujaili et~al.(2018)Al-Dujaili, Huang, Hemberg, and
  O’Reilly]{al2018adversarial}
Abdullah Al-Dujaili, Alex Huang, Erik Hemberg, and Una-May O’Reilly.
\newblock Adversarial deep learning for robust detection of binary encoded
  malware.
\newblock In \emph{2018 IEEE Security and Privacy Workshops (SPW)}, pages
  76--82. IEEE, 2018.

\bibitem[Alasmary et~al.(2020)Alasmary, Abusnaina, Jang, Abuhamad, Anwar,
  NYANG, and Mohaisen]{alasmary2020soteria}
Hisham Alasmary, Ahmed Abusnaina, Rhongho Jang, Mohammed Abuhamad, Afsah Anwar,
  D~NYANG, and David Mohaisen.
\newblock Soteria: Detecting adversarial examples in control flow graph-based
  malware classifiers.
\newblock In \emph{40th IEEE International Conference on Distributed Computing
  Systems, ICDCS}, pages 1296--1305, 2020.

\bibitem[Fleshman et~al.(2019)Fleshman, Raff, Sylvester, Forsyth, and
  McLean]{fleshman2018non}
William Fleshman, Edward Raff, Jared Sylvester, Steven Forsyth, and Mark
  McLean.
\newblock Non-negative networks against adversarial attacks.
\newblock \emph{AAAI workshop}, 2019.

\bibitem[Ceschin et~al.(2019)Ceschin, Botacin, Gomes, Oliveira, and
  Gr{\'e}gio]{ceschin2019shallow}
Fabr{\'\i}cio Ceschin, Marcus Botacin, Heitor~Murilo Gomes, Luiz~S Oliveira,
  and Andr{\'e} Gr{\'e}gio.
\newblock Shallow security: On the creation of adversarial variants to evade
  machine learning-based malware detectors.
\newblock In \emph{Proceedings of the 3rd Reversing and Offensive-oriented
  Trends Symposium}, pages 1--9, 2019.

\bibitem[Demetrio et~al.(2021{\natexlab{b}})Demetrio, Biggio, Lagorio, Roli,
  and Armando]{demetrio2021functionality}
Luca Demetrio, Battista Biggio, Giovanni Lagorio, Fabio Roli, and Alessandro
  Armando.
\newblock Functionality-preserving black-box optimization of adversarial
  windows malware.
\newblock \emph{IEEE Transactions on Information Forensics and Security},
  16:\penalty0 3469--3478, 2021{\natexlab{b}}.

\bibitem[Errica et~al.(2019)Errica, Podda, Bacciu, and Micheli]{Errica2020fair}
Federico Errica, Marco Podda, Davide Bacciu, and Alessio Micheli.
\newblock A fair comparison of graph neural networks for graph classification.
\newblock \emph{arXiv preprint arXiv:1912.09893}, 2019.

\bibitem[Xu et~al.(2019)Xu, Hu, Leskovec, and Jegelka]{xu2018how}
Keyulu Xu, Weihua Hu, Jure Leskovec, and Stefanie Jegelka.
\newblock How powerful are graph neural networks?
\newblock In \emph{International Conference on Learning Representations}, 2019.

\bibitem[Raff et~al.(2018)Raff, Barker, Sylvester, Brandon, Catanzaro, and
  Nicholas]{raff2018malware}
Edward Raff, Jon Barker, Jared Sylvester, Robert Brandon, Bryan Catanzaro, and
  Charles~K Nicholas.
\newblock Malware detection by eating a whole exe.
\newblock In \emph{Workshops at the Thirty-Second AAAI Conference on Artificial
  Intelligence}, 2018.

\bibitem[Jiang et~al.(2019)Jiang, Turki, and Wang]{Jiang2019}
Haodi Jiang, Turki Turki, and Jason~T.L. Wang.
\newblock {DLGraph: Malware Detection Using Deep Learning and Graph Embedding}.
\newblock \emph{Proceedings - 17th IEEE International Conference on Machine
  Learning and Applications, ICMLA 2018}, pages 1029--1033, 2019.
\newblock \doi{10.1109/ICMLA.2018.00168}.

\bibitem[John et~al.()John, Thomas, and Emmanuel]{john2020graph}
Teenu~S John, Tony Thomas, and Sabu Emmanuel.
\newblock Graph convolutional networks for android malware detection with
  system call graphs.
\newblock In \emph{2020 Third ISEA Conference on Security and Privacy
  (ISEA-ISAP)}, pages 162--170. IEEE.

\bibitem[Pei et~al.(2020)Pei, Yu, and Tian]{pei2020amalnet}
Xinjun Pei, Long Yu, and Shengwei Tian.
\newblock Amalnet: A deep learning framework based on graph convolutional
  networks for malware detection.
\newblock \emph{Computers and Security}, page 101792, 2020.

\bibitem[Yan et~al.(2019)Yan, Yan, and Jin]{yan2019classifying}
Jiaqi Yan, Guanhua Yan, and Dong Jin.
\newblock Classifying malware represented as control flow graphs using deep
  graph convolutional neural network.
\newblock In \emph{2019 49th Annual IEEE/IFIP International Conference on
  Dependable Systems and Networks (DSN)}, pages 52--63. IEEE, 2019.

\bibitem[W{\"u}chner et~al.(2015)W{\"u}chner, Ochoa, and
  Pretschner]{wuchner2015robust}
Tobias W{\"u}chner, Mart{\'\i}n Ochoa, and Alexander Pretschner.
\newblock Robust and effective malware detection through quantitative data flow
  graph metrics.
\newblock In \emph{International Conference on Detection of Intrusions and
  Malware, and Vulnerability Assessment}, pages 98--118. Springer, 2015.

\bibitem[Hashemi et~al.(2017)Hashemi, Azmoodeh, Hamzeh, and
  Hashemi]{hashemi2017graph}
Hashem Hashemi, Amin Azmoodeh, Ali Hamzeh, and Sattar Hashemi.
\newblock Graph embedding as a new approach for unknown malware detection.
\newblock \emph{Journal of Computer Virology and Hacking Techniques},
  13\penalty0 (3):\penalty0 153--166, 2017.

\bibitem[Nguyen et~al.(2018)Nguyen, Le~Nguyen, Nguyen, and
  Quan]{nguyen2018auto}
Minh~Hai Nguyen, Dung Le~Nguyen, Xuan~Mao Nguyen, and Tho~Thanh Quan.
\newblock Auto-detection of sophisticated malware using lazy-binding control
  flow graph and deep learning.
\newblock \emph{Computers and Security}, 76:\penalty0 128--155, 2018.

\bibitem[Frenklach et~al.(2021)Frenklach, Cohen, Shabtai, and
  Puzis]{frenklach2021android}
Tatiana Frenklach, Dvir Cohen, Asaf Shabtai, and Rami Puzis.
\newblock Android malware detection via an app similarity graph.
\newblock \emph{Computers and Security}, 109:\penalty0 102386, 2021.

\bibitem[Ou and Xu(2022)]{ou2022s3feature}
Fan Ou and Jian Xu.
\newblock S3feature: A static sensitive subgraph-based feature for android
  malware detection.
\newblock \emph{Computers and Security}, 112:\penalty0 102513, 2022.

\bibitem[Schranko~de Oliveira and Sassi(2019)]{schranko2019behavioral}
Angelo Schranko~de Oliveira and Renato~Jos{\'e} Sassi.
\newblock Behavioral malware detection using deep graph convolutional neural
  networks.
\newblock 2019.

\bibitem[Gao et~al.(2021)Gao, Cheng, and Zhang]{gao2021gdroid}
Han Gao, Shaoyin Cheng, and Weiming Zhang.
\newblock Gdroid: Android malware detection and classification with graph
  convolutional network.
\newblock \emph{Computers and Security}, 106:\penalty0 102264, 2021.

\bibitem[Z{\"u}gner and G{\"u}nnemann(2019)]{zugner2019adversarial}
Daniel Z{\"u}gner and Stephan G{\"u}nnemann.
\newblock Adversarial attacks on graph neural networks via meta learning.
\newblock \emph{arXiv preprint arXiv:1902.08412}, 2019.

\bibitem[Grosse et~al.(2017)Grosse, Papernot, Manoharan, Backes, and
  McDaniel]{grosse2017adversarial}
Kathrin Grosse, Nicolas Papernot, Praveen Manoharan, Michael Backes, and
  Patrick McDaniel.
\newblock Adversarial examples for malware detection.
\newblock In \emph{European Symposium on Research in Computer Security}, pages
  62--79. Springer, 2017.

\bibitem[Kreuk et~al.(2018)Kreuk, Barak, Aviv-Reuven, Baruch, Pinkas, and
  Keshet]{kreuk2018deceiving}
Felix Kreuk, Assi Barak, Shir Aviv-Reuven, Moran Baruch, Benny Pinkas, and
  Joseph Keshet.
\newblock Deceiving end-to-end deep learning malware detectors using
  adversarial examples.
\newblock \emph{arXiv preprint arXiv:1802.04528}, 2018.

\bibitem[Rigaki and Garcia(2018)]{rigaki2018bringing}
Maria Rigaki and Sebastian Garcia.
\newblock Bringing a gan to a knife-fight: Adapting malware communication to
  avoid detection.
\newblock In \emph{2018 IEEE Security and Privacy Workshops (SPW)}, pages
  70--75. IEEE, 2018.

\bibitem[Kawai et~al.(2019)Kawai, Ota, and Dong]{kawai2019improved}
Masataka Kawai, Kaoru Ota, and Mianxing Dong.
\newblock Improved malgan: Avoiding malware detector by leaning cleanware
  features.
\newblock In \emph{2019 International Conference on Artificial Intelligence in
  Information and Communication (ICAIIC)}, pages 040--045. IEEE, 2019.

\bibitem[Vaya and Sen(2020)]{pesidious}
Chandni Vaya and IBM~Security Sen, Bedang.
\newblock Malware mutation using deep reinforcement learning and gan.
\newblock \emph{Hack in the Box}, 2020.

\bibitem[Abusnaina et~al.(2019)Abusnaina, Khormali, Alasmary, Park, Anwar, and
  Mohaisen]{abusnaina2019adversarial}
Ahmed Abusnaina, Aminollah Khormali, Hisham Alasmary, Jeman Park, Afsah Anwar,
  and Aziz Mohaisen.
\newblock Adversarial learning attacks on graph-based iot malware detection
  systems.
\newblock In \emph{2019 IEEE 39th International Conference on Distributed
  Computing Systems (ICDCS)}, pages 1296--1305. IEEE, 2019.

\bibitem[Zhang et~al.(2019)Zhang, Chen, Song, Boning, inderjit dhillon, and
  Hsieh]{zhang2018the}
Huan Zhang, Hongge Chen, Zhao Song, Duane Boning, inderjit dhillon, and Cho-Jui
  Hsieh.
\newblock The limitations of adversarial training and the blind-spot attack.
\newblock In \emph{International Conference on Learning Representations}, 2019.

\bibitem[Sadeghzadeh et~al.(2021)Sadeghzadeh, Tajali, and Jalili]{9408630}
Amir~Mahdi Sadeghzadeh, Behrad Tajali, and Rasool Jalili.
\newblock Awa: Adversarial website adaptation.
\newblock \emph{IEEE Transactions on Information Forensics and Security},
  16:\penalty0 3109--3122, 2021.
\newblock \doi{10.1109/TIFS.2021.3074295}.

\bibitem[Rathore et~al.(2020)Rathore, Sahay, Nikam, and
  Sewak]{rathore2020robust}
Hemant Rathore, Sanjay~K Sahay, Piyush Nikam, and Mohit Sewak.
\newblock Robust android malware detection system against adversarial attacks
  using q-learning.
\newblock \emph{Information Systems Frontiers}, pages 1--16, 2020.

\bibitem[Khoda et~al.(2019)Khoda, Imam, Kamruzzaman, Gondal, and
  Rahman]{khoda2019robust}
Mahbub~E Khoda, Tasadduq Imam, Joarder Kamruzzaman, Iqbal Gondal, and Ashfaqur
  Rahman.
\newblock Robust malware defense in industrial iot applications using machine
  learning with selective adversarial samples.
\newblock \emph{IEEE Transactions on Industry Applications}, 2019.

\bibitem[Wu et~al.(2018)Wu, Shi, Yang, and Li]{wu2018enhancing}
Cangshuai Wu, Jiangyong Shi, Yuexiang Yang, and Wenhua Li.
\newblock Enhancing machine learning based malware detection model by
  reinforcement learning.
\newblock In \emph{Proceedings of the 8th International Conference on
  Communication and Network Security}, pages 74--78, 2018.

\bibitem[Chen et~al.(2019)Chen, Ren, Yu, Hussain, and Liu]{chen2019adversarial}
Bingcai Chen, Zhongru Ren, Chao Yu, Iftikhar Hussain, and Jintao Liu.
\newblock Adversarial examples for cnn-based malware detectors.
\newblock \emph{IEEE Access}, 7:\penalty0 54360--54371, 2019.

\bibitem[Li et~al.(2020)Li, Li, Ye, and Xu]{li2020enhancing}
Deqiang Li, Qianmu Li, Yanfang Ye, and Shouhuai Xu.
\newblock Enhancing deep neural networks against adversarial malware examples.
\newblock \emph{arXiv preprint arXiv:2004.07919}, 2020.

\bibitem[Demontis et~al.(2017)Demontis, Melis, Biggio, Maiorca, Arp, Rieck,
  Corona, Giacinto, and Roli]{demontis2017yes}
Ambra Demontis, Marco Melis, Battista Biggio, Davide Maiorca, Daniel Arp,
  Konrad Rieck, Igino Corona, Giorgio Giacinto, and Fabio Roli.
\newblock Yes, machine learning can be more secure! a case study on android
  malware detection.
\newblock \emph{IEEE Transactions on Dependable and Secure Computing},
  16\penalty0 (4):\penalty0 711--724, 2017.

\bibitem[Kumar et~al.(2018)Kumar, Xiaosong, Khan, Kumar, and
  Ahad]{arp2014drebin}
Rajesh Kumar, Zhang Xiaosong, Riaz~Ullah Khan, Jay Kumar, and Ijaz Ahad.
\newblock Effective and explainable detection of android malware based on
  machine learning algorithms.
\newblock In \emph{Proceedings of the 2018 International Conference on
  Computing and Artificial Intelligence}, pages 35--40, 2018.

\bibitem[Ebrahimi et~al.(2021)Ebrahimi, Zhang, Hu, Raza, and
  Chen]{Ebrahimi2020Binary}
Mohammadreza Ebrahimi, Ning Zhang, James~Lee Hu, Muhammad~Taqi Raza, and
  Hsinchun Chen.
\newblock Binary black-box evasion attacks against deep learning-based static
  malware detectors with adversarial byte-level language model.
\newblock In \emph{AAAI Conference on Artificial Intelligence, Workshop on
  Robust, Secure, and Efficient Machine Learning (RSEML)}, 2021.

\bibitem[Lucas et~al.(2021)Lucas, Sharif, Bauer, Reiter, and
  Shintre]{10.1145/3433210.3453086}
Keane Lucas, Mahmood Sharif, Lujo Bauer, Michael~K. Reiter, and Saurabh
  Shintre.
\newblock Malware makeover: Breaking ml-based static analysis by modifying
  executable bytes.
\newblock In \emph{Proceedings of the 2021 ACM Asia Conference on Computer and
  Communications Security}, page 744–758, 2021.

\bibitem[IDA()]{IDA}
Ida pro.
\newblock \emph{https://www.hex-rays.com.}
\newblock [Online; accessed 11-Jun-2021].

\bibitem[IAT()]{IAT}
Portable executable format.
\newblock
  \emph{https://docs.microsoft.com/en-us/windows/win32/debug/pe-format.}
\newblock [Online; accessed 11-Jun-2021].

\bibitem[dot()]{dotnet}
Anatomy of a .net assembly – methods.
\newblock
  \emph{https://www.red-gate.com/simple-talk/blogs/anatomy-of-a-net-assembly-methods.}
\newblock [Online; accessed 11-Jun-2021].

\bibitem[Guilfanov(1997)]{guilfanov1997fast}
Ilfak Guilfanov.
\newblock Fast library identification and recognition technology.
\newblock \emph{Li{\`e}ge, Belgium: DataRescue}, 1997.

\bibitem[Manning and Schutze(1999)]{manning1999foundations}
Christopher Manning and Hinrich Schutze.
\newblock \emph{Foundations of statistical natural language processing}.
\newblock MIT press, 1999.

\bibitem[vir({\natexlab{a}})]{virusshare}
{Virusshare}.
\newblock \emph{{https://www.virusshare.com}}, {\natexlab{a}}.
\newblock [Online; accessed 11-Jun-2021].

\bibitem[vir({\natexlab{b}})]{virustotal}
{VirusTotal Intelligence Service}.
\newblock \emph{{https://www.virustotal.com}}, {\natexlab{b}}.
\newblock [Online; accessed 11-Jun-2021].

\bibitem[Sebasti{\'a}n et~al.(2016)Sebasti{\'a}n, Rivera, Kotzias, and
  Caballero]{sebastian2016avclass}
Marcos Sebasti{\'a}n, Richard Rivera, Platon Kotzias, and Juan Caballero.
\newblock Avclass: A tool for massive malware labeling.
\newblock In \emph{International symposium on research in attacks, intrusions,
  and defenses}, pages 230--253. Springer, 2016.

\bibitem[fre()]{freewarefiles}
{Freeware Files}.
\newblock \emph{{https://www.freewarefiles.com}}.
\newblock [Online; accessed 11-Jun-2021].

\bibitem[Inn()]{Inno}
Inno setup.
\newblock \emph{https://jrsoftware.org/isinfo.php.}
\newblock [Online; accessed 11-Jun-2021].

\bibitem[uni()]{uniextractor}
{UniExtractor}.
\newblock \emph{{https://github.com/Bioruebe/UniExtract2}}.
\newblock [Online; accessed 11-Jun-2021].

\bibitem[lim()]{lime}
{Lime Crypter}.
\newblock \emph{{https://github.com/NYAN-x-CAT/Lime-Crypter}}.
\newblock [Online; accessed 11-Jun-2021].

\bibitem[Yan et~al.(2008)Yan, Zhang, and Ansari]{yan2008revealing}
Wei Yan, Zheng Zhang, and Nirwan Ansari.
\newblock Revealing packed malware.
\newblock \emph{IEEE Security and PrivaCy}, 6\penalty0 (5):\penalty0 65--69,
  2008.

\bibitem[Holt(2016)]{holt2016cybercrime}
Thomas~J Holt.
\newblock \emph{Cybercrime through an interdisciplinary lens}.
\newblock Taylor and Francis, 2016.

\bibitem[Balci and Tester(2016)]{balci2016art}
Ege Balci and Penetration Tester.
\newblock Art of anti detection--1 introduction to av and detection techniques.
\newblock 2016.

\bibitem[Rac()]{Raccoon1}
{Hunting Raccoon Stealer: The New Masked Bandit on the Block}.
\newblock
  \emph{{https://www.cybereason.com/blog/hunting-raccoon-stealer-the-new-masked-bandit-on-the-block}}.
\newblock [Online; accessed 11-Jun-2021].

\bibitem[Beek et~al.(2018)Beek, Diwakar, Yashashree, German, Niamh, Francisca,
  Eric, Thomas, et~al.]{beek2017mcafee}
C~Beek, D~Diwakar, G~Yashashree, L~German, M~Niamh, M~Francisca, P~Eric,
  R~Thomas, et~al.
\newblock Mcafee labs threats report-june 2017.
\newblock 2018.

\bibitem[die()]{die}
{Detect it easy}.
\newblock \emph{{https://github.com/horsicq/Detect-It-Easy}}.
\newblock [Online; accessed 11-Jun-2021].

\bibitem[Wang et~al.(2019)Wang, Yu, Zheng, Gan, Gai, Ye, Li, Zhou, Huang, Ma,
  et~al.]{wang2019deep}
Minjie Wang, Lingfan Yu, Da~Zheng, Quan Gan, Yu~Gai, Zihao Ye, Mufei Li,
  Jinjing Zhou, Qi~Huang, Chao Ma, et~al.
\newblock Deep graph library: Towards efficient and scalable deep learning on
  graphs.
\newblock 2019.

\bibitem[Swinnen and Mesbahi(2014)]{swinnen2014one}
Arne Swinnen and Alaeddine Mesbahi.
\newblock One packer to rule them all: Empirical identification, comparison and
  circumvention of current antivirus detection techniques.
\newblock \emph{BlackHat USA}, 2014.

\bibitem[Snow and Monrose(2012)]{snow2012automatic}
Kevin~Z Snow and Fabian Monrose.
\newblock Automatic hooking for forensic analysis of document-based code
  injection attacks.
\newblock In \emph{European Workshop on System Security}. Citeseer, 2012.

\bibitem[Green et~al.(2015)Green, Chandnani, and
  Christensen]{green2015detecting}
Jonathon~Patrick Green, Anjali~Doulatram Chandnani, and Simon~David
  Christensen.
\newblock Detecting script-based malware using emulation and heuristics.
\newblock March~31 2015.
\newblock US Patent 8,997,233.

\bibitem[Golshan and Binder(2016)]{golshan2016systems}
Ali Golshan and James~S Binder.
\newblock Systems and methods for virtualization and emulation assisted malware
  detection.
\newblock December~13 2016.
\newblock US Patent 9,519,781.

\bibitem[Kang et~al.(2009)Kang, Yin, Hanna, McCamant, and
  Song]{kang2009emulating}
Min~Gyung Kang, Heng Yin, Steve Hanna, Stephen McCamant, and Dawn Song.
\newblock Emulating emulation-resistant malware.
\newblock In \emph{Proceedings of the 1st ACM workshop on Virtual machine
  security}, pages 11--22, 2009.

\end{thebibliography}


\end{document}